\documentclass[preprint,12pt]{elsarticle}

\usepackage{amssymb}

\usepackage{lineno}

\journal{Computer Science Review}

\usepackage{rotating}

\usepackage{lmodern}
\usepackage{array}
\usepackage{graphicx}
\usepackage{multirow}
\usepackage{tabularx}
\usepackage{lscape}

\usepackage{rotating}
\usepackage{multirow}
\usepackage{fullpage}
\usepackage{anysize}
\marginsize{2cm}{2cm}{2cm}{2cm}
\usepackage{filecontents}

\usepackage[boxed,ruled,vlined,linesnumbered]{algorithm2e}
\usepackage{url}
\usepackage{framed}

\newcommand*\wrapletters[1]{\wr@pletters#1\@nil}
\def\wr@pletters#1#2\@nil{#1\allowbreak\if&#2&\else\wr@pletters#2\@nil\fi}
\usepackage{todonotes}
\usepackage{tikz}
\usetikzlibrary{automata,matrix,shapes,arrows,positioning,chains,calc}
\usetikzlibrary{snakes}
\usetikzlibrary{arrows,scopes}
\usetikzlibrary{positioning,chains,fit,shapes,calc}
\usepackage{caption}
\usepackage{subcaption}
\usepackage[none]{hyphenat}
\usepackage{times}
\usepackage{amsmath}
\SetKwHangingKw{Local}{Local}
\SetKwHangingKw{Global}{Global}
\SetKwHangingKw{Constant}{Constant}
\SetKwHangingKw{Input}{Input}
\SetKwHangingKw{Alias}{Alias}
\SetKwHangingKw{Output}{Output}
\SetKwHangingKw{SideEffect}{Side effects}
\SetKwHangingKw{InternalEvent}{Internal event}
\SetKwHangingKw{External}{External}
\SetKwHangingKw{ExternalEvent}{External event}
\SetKwHangingKw{Precondition}{Precondition :}
\SetKwHangingKw{Postcondition}{Postcondition :}
\SetKwHangingKw{Upon}{Upon}
\SetKwHangingKw{DoForever}{Do Forever}
\SetKwHangingKw{PVab}{Persistent variables:}

\usepackage{theorem}
\usepackage{wrapfig}
\usepackage{amssymb}
\usepackage{xspace}

\newcommand{\BB}{\vspace*{-\medskipamount}}
\newcommand{\BBB}{\vspace*{-\bigskipamount}}

\newcommand{\FF}{\vspace*{\medskipamount}}

\newcommand{\remove}[1]{}

\newcommand{\Subsubsubsection}[1]{\FF \noindent $\bullet$ {\bf #1}~~~~~~}
\usepackage[T1]{fontenc}
\usepackage[utf8]{inputenc}

\usepackage{setspace}

\usepackage{csquotes}
\usepackage{enumitem}

\usepackage{tablefootnote}
\usepackage{longtable}

\usepackage{multirow}
\usepackage{hhline}

\begin{document}
\title{Security and Privacy Aspects in MapReduce on Clouds: A Survey}
\begin{frontmatter}

\date{}
\author[add1]{Philip Derbeko}
  \ead{philip.derbeko@emc.com}
  \author[add2]{Shlomi Dolev\corref{cor1}}

  \ead{dolev@cs.bgu.ac.il}
  \author[add2]{Ehud Gudes}
  \ead{ehud@cs.bgu.ac.il}
  \author[add2]{Shantanu Sharma}
  \ead{sharmas@cs.bgu.ac.il}

\cortext[cor1]{Partially supported by the Rita Altura Trust Chair in Computer Sciences, Lynne and William Frankel Center for Computer Sciences, Israel Science Foundation (grant 428/11), the Israeli Internet Association, and the Ministry of Science and Technology, Infrastructure Research in the Field of Advanced Computing and Cyber Security.}

  \address[add1]{EMC, Beer-Sheva, Israel COE.}
  \address[add2]{Department of Computer Science, Ben-Gurion University of the Negev, Israel.}

\begin{abstract}
MapReduce is a programming system for distributed processing large-scale data in an efficient and fault tolerant manner on a private, public, or hybrid cloud. MapReduce is extensively used daily around the world as an efficient distributed computation tool for a large class of problems, \textit{e}.\textit{g}., search, clustering, log analysis, different types of join operations, matrix multiplication, pattern matching, and analysis of social networks. Security and privacy of data and MapReduce computations are essential concerns when a MapReduce computation is executed in public or hybrid clouds. In order to execute a MapReduce job in public and hybrid clouds, authentication of mappers-reducers, confidentiality of data-computations, integrity of data-computations, and correctness-freshness of the outputs are required. Satisfying these requirements shield the operation from several types of attacks on data and MapReduce computations. In this paper, we investigate and discuss security and privacy challenges and requirements, considering a variety of adversarial capabilities, and characteristics in the scope of MapReduce. We also provide a review of existing security and privacy protocols for MapReduce and discuss their overhead issues.
\end{abstract}

\end{frontmatter}
\newpage
\tableofcontents


\pagenumbering{arabic}
\setcounter{page}{1}
\section{Introduction}
\label{sec:introduction}
Cloud computing~\cite{mell2011nist} infrastructure provides on-demand, easy, and scalable access to a shared pool of configurable resources, without worrying about managing those resources. Details about cloud computing can be found in~\cite{buyya2010cloud,zhang2010cloud}. Clouds provide three types of services, as follows: (\textit{i}) \emph{infrastructure-as-a-service}, IaaS, provides infrastructure in terms of virtual machines, storage, and networks, (\textit{ii}) \emph{platform-as-a-service}, PaaS, provides a scalable software platform allowing the development of custom applications, and (\textit{iii}) \emph{software-as-a-service}, SaaS, provides software running in clouds as a service, for example, emails and databases. Clouds can be classified into three types, as follows: (\textit{i}) \textit{public cloud}: a cloud that provides services to many users and is not under the control of a single exclusive user, (\textit{ii}) \textit{private cloud}: a cloud that has its proprietary resources and is under the control of a single exclusive user, and (\textit{iii}) \textit{hybrid cloud}: a combination of public and private clouds.

One of the most common \emph{platform-as-a-service} computational paradigms is MapReduce~\cite{DBLP:conf/osdi/DeanG04}, introduced by Google in 2004. MapReduce provides an efficient and fault tolerant parallel processing of large-scale data without any costly and dedicated computing node like a supercomputer.
At the beginning, MapReduce was designed to be deployed on-premises under mistaken assumption that local environment can be completely trusted. Thus, security and privacy aspects were overlooked in the initial design.
As MapReduce gained popularity the lack of security and privacy in on-premises deployment become severe shortcoming. In addition, MapReduce is being deployed on both hybrid and public clouds, which are prone to many attacks and security threats. In the current days, several public clouds, \textit{e}.\textit{g}., Amazon Elastic MapReduce, Google App Engine, IBM's Blue Cloud, and Microsoft Azure, enable users to perform MapReduce cloud computations without considering physical infrastructures and software installations. Thus, the deployment of MapReduce in public clouds enables users to process large-scale data in a cost-effective manner and establishes a relationship between two independent entities, \textit{i}.\textit{e}., clouds and MapReduce. As a downside, the deployment of MapReduce in hybrid and public cloud needs to deal with many attacks on the communication networks and (the three service layers of) the cloud.

\begin{sidewaysfigure}
\centering
\includegraphics[scale=0.6]{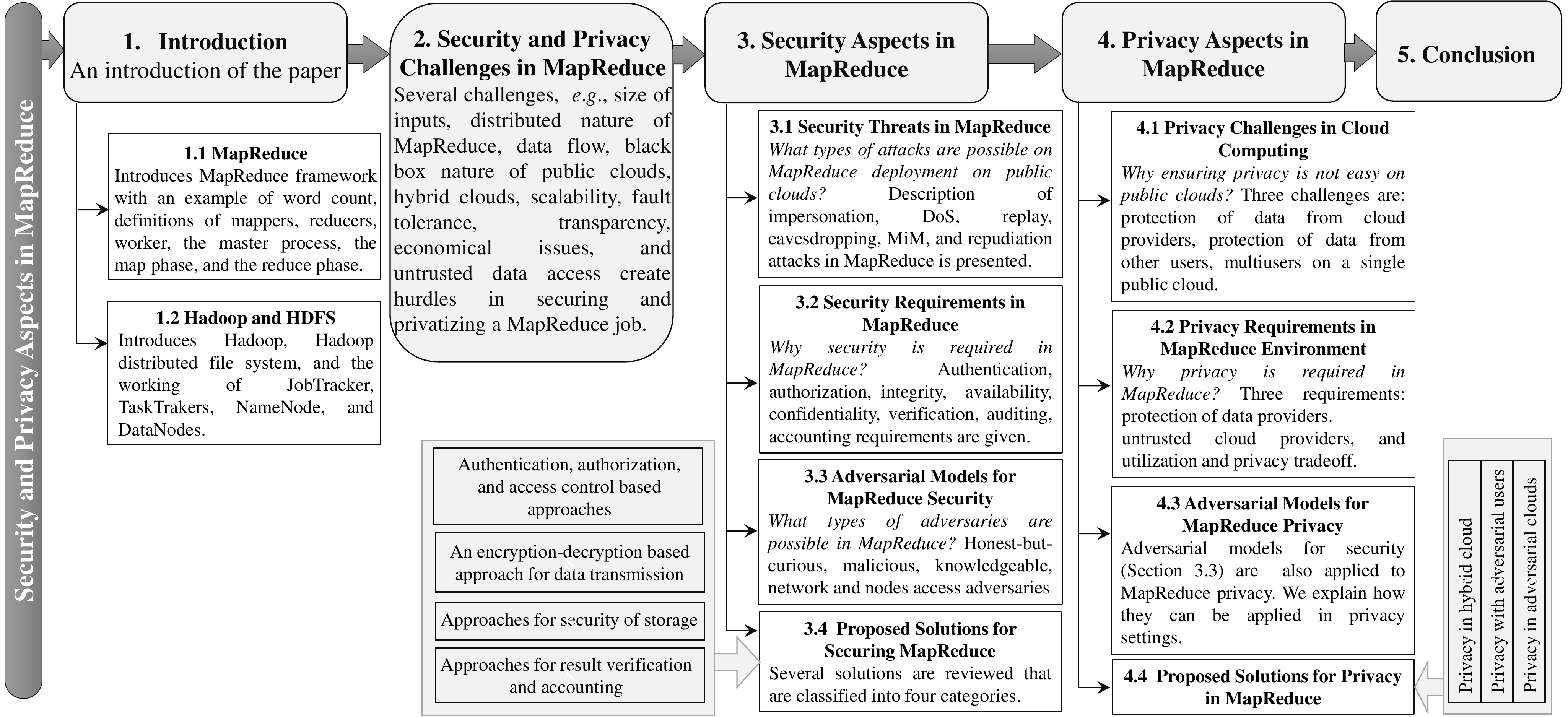}
\caption{Schematic map of the paper.}
\label{fig:Outline of the paper}
\end{sidewaysfigure}


Data processing in the cloud highlights a tradeoff between the ease of processing and security-privacy of data and computations. Specifically, on one hand, the deployment of MapReduce in a well-managed public cloud provides economical and carefree resource management. On the other hand, public clouds do not guarantee the rigorous security and privacy of computations as well as stored data. Private clouds provide security and privacy of data as well as computations, due to users' ability to physically and electronically constrain data access and execution of computations. However, the user of the private cloud manages the nodes, updates software, and replaces the failed nodes. Such management is time consuming and incurs huge monetary cost. 

Our focus is on the security and privacy issues of MapReduce environment in public or hybrid clouds. Private cloud environments are more secure due to a physical security of the cloud. Many of the reviewed below results are applicable to both public and hybrid clouds, unless stated otherwise (for instance, see hybrid cloud specific research in~\cite{DBLP:conf/ccs/ZhangZCWR11, 13middleware} and Section~\ref{subsubsec:Data privacy in hybrid clouds}).
Even though there is a plethora of additional projects and frameworks that add functionality on top of MapReduce (see Apache Hive~\cite{DBLP:journals/pvldb/ThusooSJSCALWM09}, Cloudera Impala\footnote{http://impala.io/}, HBase~\cite{george2011hbase}, Apache Zookeeper\footnote{https://zookeeper.apache.org/.}, Thrift\footnote{https://thrift.apache.org/.}, and Apache Solr\footnote{http://lucene.apache.org/solr/}), this paper only reviews security related projects in Section~\ref{sec:Challenges in MapReduce Environment and Adversary Models} (readers interested in security and privacy issues of other projects may refer to~\cite{DBLP:journals/csur/SakrLF13}).

Security aspects in the context of MapReduce are crucial in order to authenticate and authorize users, auditing-confidentiality-integrity of data and computation, availability of mappers and reducers, and verification of outputs. Security of MapReduce ensures a \textit{legitimate} functionality of the framework. A secure MapReduce framework deals with the following attacks: attacks on authentication (impersonation and replay attacks), attacks on confidentiality (eavesdropping and man-in-the-middle attacks), data tampering (modification of input data, intermediate outputs, and the final outputs), hardware tampering, software tampering (modification of mappers and reducers), denial-of-service, interception-release of data as well as computations, and communication analysis.

On the contrary, privacy aspects assume \textit{legitimate} functionality of the framework and thus, are built on top of security. On top of the correctly functioning framework, privacy in the context of MapReduce is an ability of each participating party (data providers, cloud providers, and users) to prevent other, possibly adversarial parties from observing data, codes, computations, and outputs. In order to ensure privacy, a MapReduce algorithm in public clouds hides data storage as well as the computation to public clouds and adversarial users. Additional distinction between security and privacy is that security is much more of a binary issue, \textit{i}.\textit{e}., either the attack succeeds or not, whereas in a privacy setting there is a tradeoff between privacy of the data and utilization of the framework.

\medskip \noindent \textbf{Scope and outline of the paper.} In this paper, we discuss the challenges, requirements, adversarial models, attack scenarios and proposed solutions to the security and privacy concerns in the context of MapReduce (see Figure~\ref{fig:Outline of the paper}). Overview of MapReduce environment and its open-source software framework, Apache Hadoop, are given in Section~\ref{subsec:MapReduce} and Section~\ref{subsec:Hadoop}, respectively. In Section~\ref{sec:Challenges in MapReduce Environment and Adversary Models}, we discuss security challenges involved in a MapReduce computation.

Security aspects of MapReduce are presented in Section~\ref{sec:Security Aspects in MapReduce}, where we present security threats in Section~\ref{subsec:Security Threats in MapReduce}, requirements of security in MapReduce computations in Section~\ref{subsec:Security Requirements in MapReduce}, adversarial models in Section~\ref{sbsec:Adversary Models for MapReduce Security}, and some proposed solutions for security in MapReduce in Section~\ref{subsec:Proposed Solutions for Security in MapReduce}. Privacy aspects in MapReduce are presented in Section~\ref{section:Privacy Aspects in MapReduce}. We first outline privacy aspects in clouds (Section~\ref{subsection:Privacy Threats in MapReduce Computing}) and privacy requirements in MapReduce in Section~\ref{sebsec:Privacy Requirements in MapReduce}. Then, we present the common adversarial models in the context of privacy in MapReduce in Section~\ref{subsection:Adversary Models for MapReduce Privacy}, and some proposed solutions to privacy in MapReduce in Section~\ref{subsec:Proposed Solutions for Privacy in MapReduce}. We conclude and outline important research issues in Section~\ref{sec:Conclusion}.

We would like to emphasize here that we focus on the security and privacy issues of MapReduce framework.
Despite advantages of MapReduce deployments in public clouds, they also bring new hurdles in the form of security, data privacy, and computation privacy. Even though the new challenges are mainly due to the public nature of the cloud, ownership separation of platform and data, location of the service, and etc., those challenges are different from security and privacy challenges of a general cloud computing.
General security issues in the cloud such as: security of virtual machines and hypervisors, security of services and Service-Level Agreement (SLA), regulations and organization policies, and general availability in the cloud are not discussed in detail, and the interested reader may refer to~\cite{ArmbrustFGJKKLPRSZ10,Anthes10e,TakabiJA10,ZissisL12,xiao2013security,DBLP:journals/compsec/RahmanC15,DBLP:journals/isci/AliKV15} for security and privacy in the cloud. Also note that we do not consider security and privacy of MapReduce scheduling algorithms, rather than we encourage readers to have an understanding of security and privacy of mappers, reducers, and data flow.

\subsection{MapReduce Framework}
\label{subsec:MapReduce}
Parallel processing of large-scale data provides outputs in a timely manner. However, it constrains the computation due to node failure, ordering of outputs, system scalability, transparency, load balancing, fault tolerance, and synchronization among the nodes. MapReduce~\cite{DBLP:conf/osdi/DeanG04} solves these issues and executes parallel processing using a cluster of computing nodes over large-scale data, but without considering security and privacy of data and computations. Here, we provide an overview of MapReduce framework, details may be found in Chapter 2 of~\cite{leskovec2014mining}.

\begin{figure}[h]
\begin{flushleft}
\centering
\includegraphics[scale=0.5]{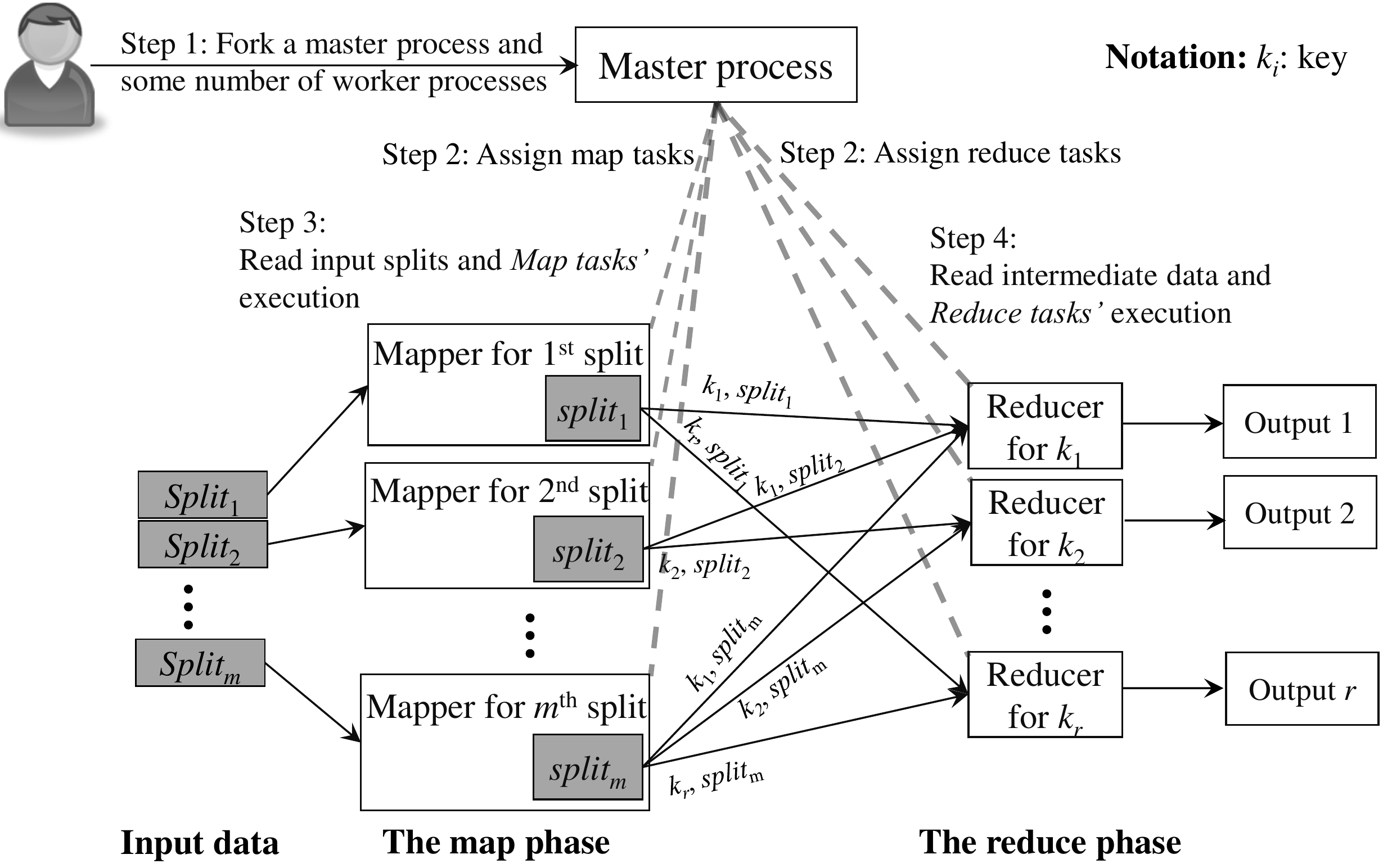}
\caption{A general execution of a MapReduce algorithm.}
\label{fig:mapreduce}
\end{flushleft}
\end{figure}

A user-defined program forks a \textit{master process} and \textit{worker processes} at different nodes; see Figure~\ref{fig:mapreduce}. The master process creates and assigns \textit{map tasks} and \textit{reduce tasks} to idle worker processes. A worker process deals with either a map task or a reduce task. The worker processes that handle the map tasks and the reduce tasks are called \textit{map workers} and \textit{reduce workers}, respectively. A MapReduce computation consists of the \emph{Map phase} and the \emph{Reduce phase}, where two user-defined functions, namely the \textit{map function} and the \textit{reduce function}, are executed over (large-scale) data, which is represented in the form of $\langle\mathit{key, value}\rangle$ pairs.

\noindent \textit{The Map Phase.} The given input data is processed in the map phase, where the map function is applied to data and produces intermediate outputs (of the form $\langle \mathit{key,value}\rangle$), where the number of bits needed to describe the $\mathit{value}$ in each $\langle\mathit{key, value}\rangle$ pair is not necessarily identical~\cite{DBLP:journals/corr/AfratiDK0U15a}. The application of the map function to a single input (for example, a tuple of a relational database or a node in a graph) is called a \textit{mapper}.

\noindent \textit{The Reduce Phase.} The Reduce phase provides the final output of MapReduce computations. The Reduce phase executes the reduce function on intermediate outputs. The application of the reduce function to a single \textit{key} and its associated list of $\mathit{value}$s is called a \textit{reducer}.

\noindent \textit{Word count example.} Word count is a traditional example to illustrate a MapReduce computation, where the task is to count the number of occurrences of each word in a collection of documents. The original input data is a collection of documents. Each mapper takes a document and implements a map function that results in a set of $\langle \mathit{key, value}\rangle$ pairs ($\{\langle w_1, 1\rangle, \langle w_2, 1\rangle, \ldots, \langle w_n, 1\rangle\}$), where each key, $w_i$, represents a word in the document, and each value is 1. The reduce task is executed subsequently, where the reduce function adds up all the values corresponding to a key. Specifically, a reducer for a key $w_i$ takes all the $\langle \mathit{key, value}\rangle$ pairs corresponding to the key (or word) $w_i$ and outputs a $\langle w_i, m\rangle$ pair, where $m$ is the total number of occurrences of the word $w_i$ in all the given documents.

\medskip \noindent\textit{Applications and models of MapReduce}. Many applications in different areas exist already for MapReduce. Among them: matrix multiplication~\cite{DBLP:journals/corr/abs-1204-1754}, similarity join~\cite{DBLP:conf/sigmod/VernicaCL10,Xiao:2008:ESJ:1367497.1367516,DBLP:conf/icde/AfratiSMPU12,Bayardo:2007:SUP:1242572.1242591}, detection of near-duplicates~\cite{DBLP:conf/www/MankuJS07}, interval join~\cite{DBLP:conf/edbt/ChawdaGNFSM14,DBLP:conf/edbt/Afrati15}, spatial join~\cite{DBLP:conf/edbt/GuptaCNFSM13,DBLP:conf/wise/GuptaC14,DBLP:conf/sigmod/TauheedHA15}, graph processing~\cite{DBLP:conf/icde/AfratiFU13,DBLP:conf/edbt/MalhotraAS14}, pattern matching~\cite{DBLP:conf/appt/LiuJCMZ09}, data cube processing~\cite{NYBR12,RP14,WGRO14,DBLP:journals/corr/Afrati0UU15}, skyline queries~\cite{DBLP:conf/icdt/AfratiKSU12}, $k$-nearest-neighbors finding~\cite{DBLP:conf/edbt/ZhangLJ12,DBLP:journals/pvldb/LuSCO12}, star-join~\cite{DBLP:conf/cloudi/ZhouZW13}, theta-join~\cite{DBLP:conf/sigmod/OkcanR11,DBLP:journals/pvldb/ZhangCW12}, and image-audio-video-graph processing~\cite{DBLP:journals/concurrency/YuWT0LV12}, are a few applications of MapReduce in the real world.
Some research models for efficient MapReduce computation are presented by Karloff et al.~\cite{DBLP:conf/soda/KarloffSV10}, Goodrich~\cite{DBLP:journals/corr/abs-1004-4708}, Lattanzi et al.~\cite{DBLP:conf/spaa/LattanziMSV11}, Pietracaprina et al.~\cite{DBLP:conf/ics/PietracaprinaPRSU12}, Goel and Munagala~\cite{DBLP:journals/corr/abs-1211-6526}, Ullman~\cite{DBLP:journals/crossroads/Ullman12}, Afrati et al.~\cite{DBLP:journals/pvldb/AfratiSSU13,DBLP:conf/ideas/AfratiU13,DBLP:journals/corr/AfratiDK0U15a,DBLP:journals/corr/AfratiD0U15}, and Fish et al.~\cite{DBLP:conf/wdag/FishKLRT15}.

\subsection{Hadoop and HDFS}
\label{subsec:Hadoop}
Apache Hadoop\footnote{http://hadoop.apache.org/} is the most known and widely used open-source software implementation of MapReduce for distributed storage and distributed processing of large-scale data on clusters of nodes. Hadoop includes three major components, as follows: (\textit{i}) Hadoop Distributed File System (HDFS)~\cite{HDFS2010}: a scalable and fault-tolerant distributed storage system, (\textit{ii}) Hadoop MapReduce, and (\textit{iii}) Hadoop Common, the common utilities, which support the other Hadoop modules. Hadoop 2.x, released in 2013, has changed the low-level architecture by separating resource management from job management (see YARN\footnote{http://hadoop.apache.org/docs/r2.6.0/hadoop-yarn/hadoop-yarn-site/YARN.html}). However, the architectural modification does not change the high-level of the described MapReduce job, and thus, is not considered here.

\begin{figure}[h]
\centering
\includegraphics[scale=0.4]{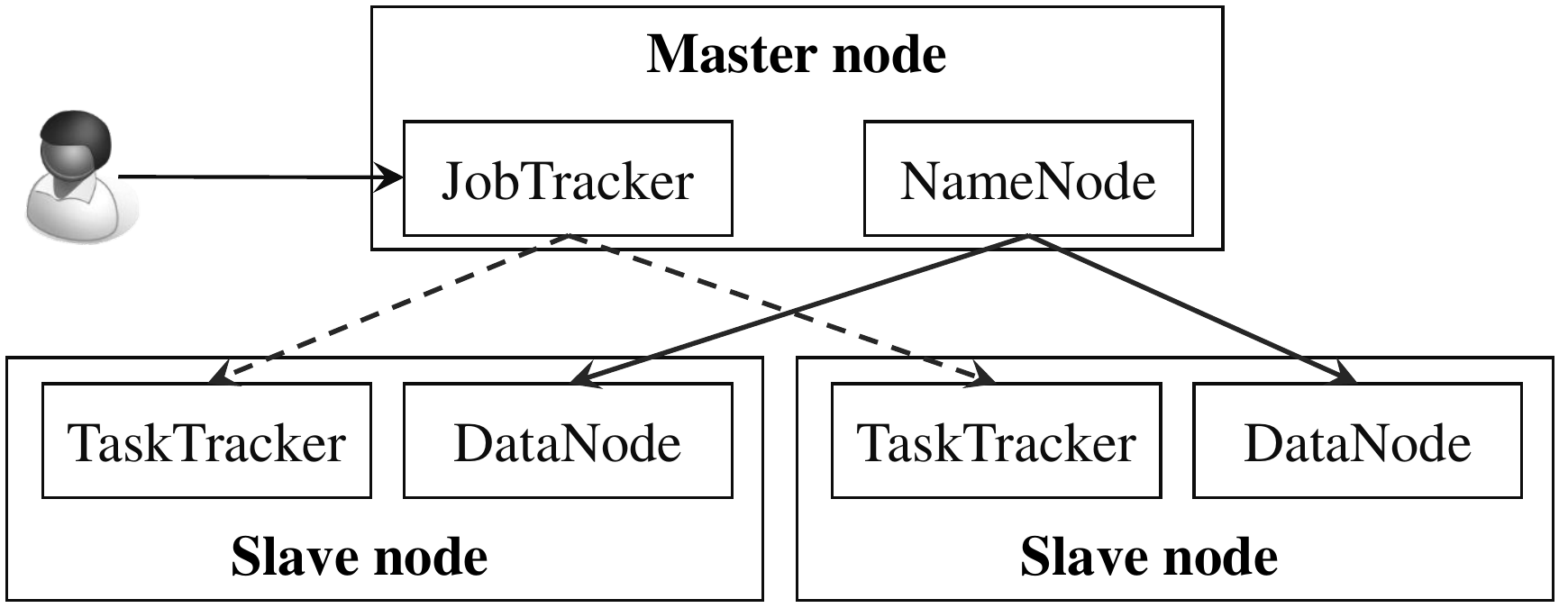}
\caption{Structure of a Hadoop cluster with one master node and two slave nodes.}
\label{fig_hadoop}
\end{figure}

Hadoop cluster consists of a master node (that runs a JobTracker and a NameNode) and several slave nodes (where each slave node runs a TaskTracker and a DataNode); see Figure~\ref{fig_hadoop}. JobTracker and TaskTrackers provide an environment for a MapReduce job execution. Specifically, JobTracker accepts a MapReduce job from a user, executes the job on (free) TaskTrackers, receives outputs from TaskTrackers, and provides outputs to the user. TaskTrackers execute assigned jobs, provide outputs to JobTracker, and periodically send heartbeat messages to JobTracker to show its presence and workload.

NameNode and DataNodes provide a distributed file system, called Hadoop Distributed File System, which supports read, write and delete operations on files, and create and delete operations on directories. NameNode manages the cluster metadata and DataNodes, which store the data. NameNode keeps information of files and directories using \emph{inodes}. Inodes store several attributes of files and directories, \textit{e}.\textit{g}., permissions, modification and access times, and disk space quotas. In HDFS, data is divided into small splits, called \emph{blocks}, (64MB and 128MB are most commonly used sizes). Each block is independently replicated at multiple DataNodes, and block replicas are processed by mappers and reducers. Each DataNode stores the corresponding block replicas, and two files in the local host's native file system are used to represent each block replica. The first file holds the data itself, and the second file holds the block's metadata. More details about Hadoop and HDFS may be found in Chapter 2 of~\cite{lin2010data}.

As the review focuses on security of MapReduce, it is fitting to provide a short overview of Hadoop and HDFS security features as well. MapReduce security is discussed later on in corresponding sections.
By default, Hadoop provides no authentication making it easy to perform destructive changes and possible attacks on other users or computing cluster.
While it is possible to configure Hadoop cluster with Kerberos authentication mechanism~\cite{kerberosurl,o2009hadoop,das2013adding}, extra work is required to do that~\cite{securehadoopurl}.
In a similar way, default configuration of HDFS provides a basic protection for the saved data by following Unix access control. However, within HDFS cluster, by default, there is no mechanism for identifying a user or group, though the user is trusted to present himself correctly. Clearly, this makes it very easy for an adversarial client to read and modify data belonging to other users.
Just like Hadoop framework, HDFS can be configured to determine user identity by its Kerberos credentials. However, just like in Hadoop's case, the configuration requires additional work and as such, most likely is not performed in all installations.

As mentioned, the job is done to provide security facilities for Hadoop (SecureMode~\cite{securehadoopurl}). The facilities include: user and service authentication (based on Kerberos mechanism), authentication for Web consoles and data confidentiality. The data confidentiality consists from features that encrypt data in-transit during Hadoop calls, including data encryption on RPC, block data transfer and HTTP access. Just like the authentication and access control mechanisms, data confidentiality requires configuration and is not configured by default.

\section{Security and Privacy Challenges in MapReduce}
\label{sec:Challenges in MapReduce Environment and Adversary Models}
The massive parallel processing style of MapReduce is substantially different from the classical computation in the cloud leading to distinct design challenges for security and privacy requirements. In this section, we present specific security and privacy design challenges for MapReduce computations in the cloud.

\medskip\noindent\textbf{Size of input data and its storage.} Input to a MapReduce job, big-data, which is described by the 4Vs: volume, velocity, variety, and veracity, implies a major challenge in securing MapReduce computations. Security and privacy techniques for processing big-data have to deal with huge amount of data, possibly arriving at high speed from different sources. Moreover, in MapReduce computations, data is partitioned into small-sized splits that are replicated and distributed to several nodes. Each split has to be transferred in a secure and private manner. This replicated and distributed nature constitutes unique challenges in terms of data storage security, as compared to a system that holds the whole data in a single place.

\medskip\noindent\textbf{Highly distributed nature of MapReduce computations.} The cloud computing itself does not necessarily imply distributed computations. However, MapReduce does require large clusters of nodes that can distributively process replicated data in parallel. The deployment of MapReduce in the cloud requires mechanisms for protecting a large-number of nodes and data, which may be in-transit and may be at the rest at the nodes. In particular, distributed processing over replicated data has a higher probability for attacks as compared to a centralized system, since attackers have a much wider range of targets to choose from. A single adversarial mapper or reducer, out of several mappers and reducers, may provide wrong outputs, copy data for future usage, modify input data, leak confidential data to a third party, or send the whole data to another user. Identifying a (single) adversarial mapper or reducer is not an easy task in the scope of MapReduce.

\medskip\noindent\textbf{Data flow.} MapReduce computations require a complex data flow among different storage nodes, different computing nodes, and different clouds, as follows:
\begin{itemize}
  \item \textit{Between data storage and computing nodes}: MapReduce computations are executed near the location of data to minimize data flow; however, ensuring an identical location of data and mappers-reducers cannot always be guaranteed. Thus, computations are typically executed at the nodes that are not storage nodes. This leads to data flow from storage to computing nodes. Moreover, some providers separate the computational cloud from the storage cloud; for example Amazon Elastic MapReduce uses two different clouds: one is for executing a MapReduce computation and the other is for storing data. This dual cloud structure requires constant data flow between the clouds. Data flow becomes more complex when organizations perform MapReduce computations in the hybrid clouds, where it is necessary for sensitive data (\textit{e}.\textit{g}., some attributes of a relation, financial data, or health records) to remain in private clouds and do not reach public clouds; while all the other data, called non-sensitive data (\textit{e}.\textit{g}., all the attributes of a relation except some attributes holding critical information), may be processed in public clouds.

  \item \textit{Between public clouds}: A MapReduce computation may be executed in more than one public cloud~\cite{DBLP:conf/dasc/ShenZYYWZ11}. In that case, data flow may occur between two master processes at two different clouds, between two mappers at two different clouds, between two reducers at two different clouds, and a mapper and a reducer at two different clouds. Such scenarios involve data flow over a public network, which is vulnerable to attacks.
\end{itemize}

\medskip\noindent\textbf{The black-box nature of public clouds supporting MapReduce.} Public clouds supporting MapReduce do not provide any information regarding the deployment, configuration and execution of mappers and reducers. PaaS infrastructure deploys and configures mappers and reducers dynamically for each computation. This opaque view of MapReduce in public clouds prohibits an efficient execution of a MapReduce computation in terms of the \emph{communication cost} (the total amount of bits that are transferred between the map phase and the reduce phase) and the replication rate~\cite{DBLP:journals/corr/AfratiDK0U15a} (the average number of key-value pairs created for each input). On the upside, the automatic resource management of public clouds offloads the user's burden. Hence, secure MapReduce computations should allow dynamic deployment and configurations of mappers and reducers in public clouds without increasing the communication cost.

\medskip\noindent\textbf{Hybrid cloud.} The hybrid cloud provides an efficient processing of sensitive and non-sensitive data. The hybrid clouds enjoy efficient and economical resource management (provided by public clouds) with security and privacy of sensitive data (provided by the private cloud). Unfortunately, MapReduce is designed to work in a single cloud, and this characteristic poses additional challenges in supporting a hybrid cloud deployment~\cite{DBLP:conf/ccs/ZhangZCWR11}. In addition, data sanitation, \textit{i}.\textit{e}., separation of sensitive and non-sensitive data, and arrangement of outputs from different clouds are additional challenges to a MapReduce computation in the hybrid clouds.

\medskip\noindent\textbf{Scalability, fault tolerance, and transparency.} MapReduce provides an efficient, scalable, distributed, and fault-free processing of replicated data in parallel. An integration of security and privacy mechanisms should not reduce efficiency, scalability, and fault tolerance of MapReduce algorithms. Also, the involvement of security and privacy protocols in MapReduce must be transparent to users, without any modification of the map and reduce functions.

\medskip\noindent\textbf{Economical issues.} An execution of MapReduce computations in public clouds is tariffed mainly for three economical factors -- data storage, the communication cost, and computation time. MapReduce algorithms also regard these three economical factors. Therefore, security and privacy mechanisms must be economically incorporated into MapReduce computations.

\medskip\noindent\textbf{Untrusted data access.} MapReduce allows great flexibility in enabling user defined computations; but at the same time, implies a great trust in users for providing mapper and reducer codes that do not impact MapReduce cluster, in terms of slowing/corrupting the entire job, modifying modifying/deleting data, and other unwanted read/write operations. Security and privacy algorithms for MapReduce should be developed to cope with corrupted or even adversarial codes, protecting data, and limiting data access of corrupted mappers and reducers.

All the above challenges to MapReduce framework in clouds indicate new security and privacy requirements. In the next sections, we discuss the requirements of security in MapReduce.


\section{Security Aspects in MapReduce}
\label{sec:Security Aspects in MapReduce}
The security of data and computations plays a significant role in MapReduce computations on both hybrid and public clouds. Without security, MapReduce computations as well as MapReduce infrastructures can be affected by several types of attacks. In this section, we present security threats and security requirements for MapReduce computations. Notice that even though some security threats and security requirements are common for MapReduce and for generic cloud computing, we will focus on security threats and security requirements in the context of MapReduce. Following that, we provide a brief summary of some existing security algorithms for MapReduce.

\subsection{Security Threats in MapReduce}
\label{subsec:Security Threats in MapReduce}
In this section, we present security threats that can harm a MapReduce computation and the framework in the absence of secure MapReduce environment. Distributed and replicated data processing in MapReduce open an opportunity for a wide range of attacks. While those attacks follow the same ideas as attacks in different cloud computation models, the exact application is different for MapReduce paradigm.

Notice that most of these attacks are specific to MapReduce deployments in public clouds, as physical security and separation of private cloud deployments significantly reduce the risk of attacks and allow a physical separation of resources from attackers.

\medskip\noindent\textbf{Impersonation attack.} \textit{Definition}: An impersonation attack occurs when an adversary successfully pretends to be a legitimate user of a system by a brute-force attack on weak passwords, weak encryption schemes, or other means.

\textit{MapReduce context}: After a successful impersonation attack an adversary can act on behalf of a legal user and can execute MapReduce jobs that may result in data leakage, data and computations tampering, or wrong computations on data~\cite{DBLP:conf/services/RuanM12,CCGrid14loganalysis,Du04uncheatablegrid}. Moreover, on public clouds under impersonation attack, an attacker may perform MapReduce computations while the impersonated user is tariffed for data storage, the communication cost and computation time.

\medskip\noindent\textbf{Denial-of-Service (DoS) attack.} \textit{Definition}: A DoS attack occurs when an adversary causes a system and the network to become non-functional and non-accessible by legitimate users.

\textit{MapReduce context}: A DoS attack occurs when an adversary makes a node, mapper, or reducer to be non-functional and non-accessible by executing undesirable and useless tasks~\cite{DBLP:conf/acsac/WeiDYG09}. Moreover, a compromised node, mapper, or reducer may result in non-functionality of other non-compromised nodes, mappers, or reducers by repeatedly sending task requests for them to execute~\cite{DBLP:conf/acsac/WeiDYG09}. In addition, if an attacker compromises enough nodes in a Hadoop cluster, it may result in the failure of the whole MapReduce framework and the network overload~\cite{6906760}.

\medskip\noindent\textbf{Replay attack.} \textit{Definition}: A replay attack occurs when an adversary resends (or replays) a captured valid message to the nodes.

\textit{MapReduce context}: A replay attack occurs when an adversary assigns some old tasks to the nodes, making them continuously busy~\cite{DBLP:conf/acsac/WeiDYG09}. In addition, an adversary can replay users' credentials to access the framework, and it may lead to impersonation~\cite{pastore2006comptia,desmedt2011relay} and DoS attacks by spinning excessive amount of nodes.

\medskip\noindent\textbf{Eavesdropping.} \textit{Definition}: An eavesdropping attack occurs when an adversary (passively) monitors the network and the nodes without consent.

\textit{MapReduce context}: An eavesdropping attack occurs when an adversary observes input data, intermediate outputs, the final outputs, and MapReduce computations without any consent from the data and computation's owner~\cite{DBLP:conf/dasc/ShenZYYWZ11,DBLP:conf/acsac/WeiDYG09,2012sapsc}.

\medskip\noindent\textbf{Man-in-the-Middle (MiM) attacks.} \textit{Definition}: In MiM, an adversary (actively) modifies, corrupts, or inserts data passing between two legitimate users of a system.

\textit{MapReduce context}: A MiM attack occurs when an adversary modifies or corrupts the computing codes, input data, intermediate outputs, or the final outputs passing between any two legitimate nodes of the framework~\cite{DBLP:conf/dasc/ShenZYYWZ11,DBLP:conf/acsac/WeiDYG09,2012sapsc}. Moreover, tampering with mappers or reducers may lead to DoS, impersonation, and replay attacks~\cite{DBLP:conf/acsac/WeiDYG09}.

\medskip\noindent\textbf{Repudiation.} \textit{Definition}: A repudiation attack occurs when a node falsely denies processing a sent message or a task execution.

\textit{MapReduce context}: A repudiation attack occurs when a mapper or reducer falsely denies an execution request that it already performed~\cite{Du04uncheatablegrid}. 

Despite the severity of all the above mentioned attacks, by default, MapReduce does not provide any way to encounter them. (Details of the above mentioned attacks on the computer networks are given in~\cite{william2006cryptography,pastore2006comptia}.) In the next section, we present security requirements in MapReduce.

\subsection{Security Requirements in MapReduce}
\label{subsec:Security Requirements in MapReduce}
\begin{figure}[h]
\centering
\includegraphics[scale=0.45]{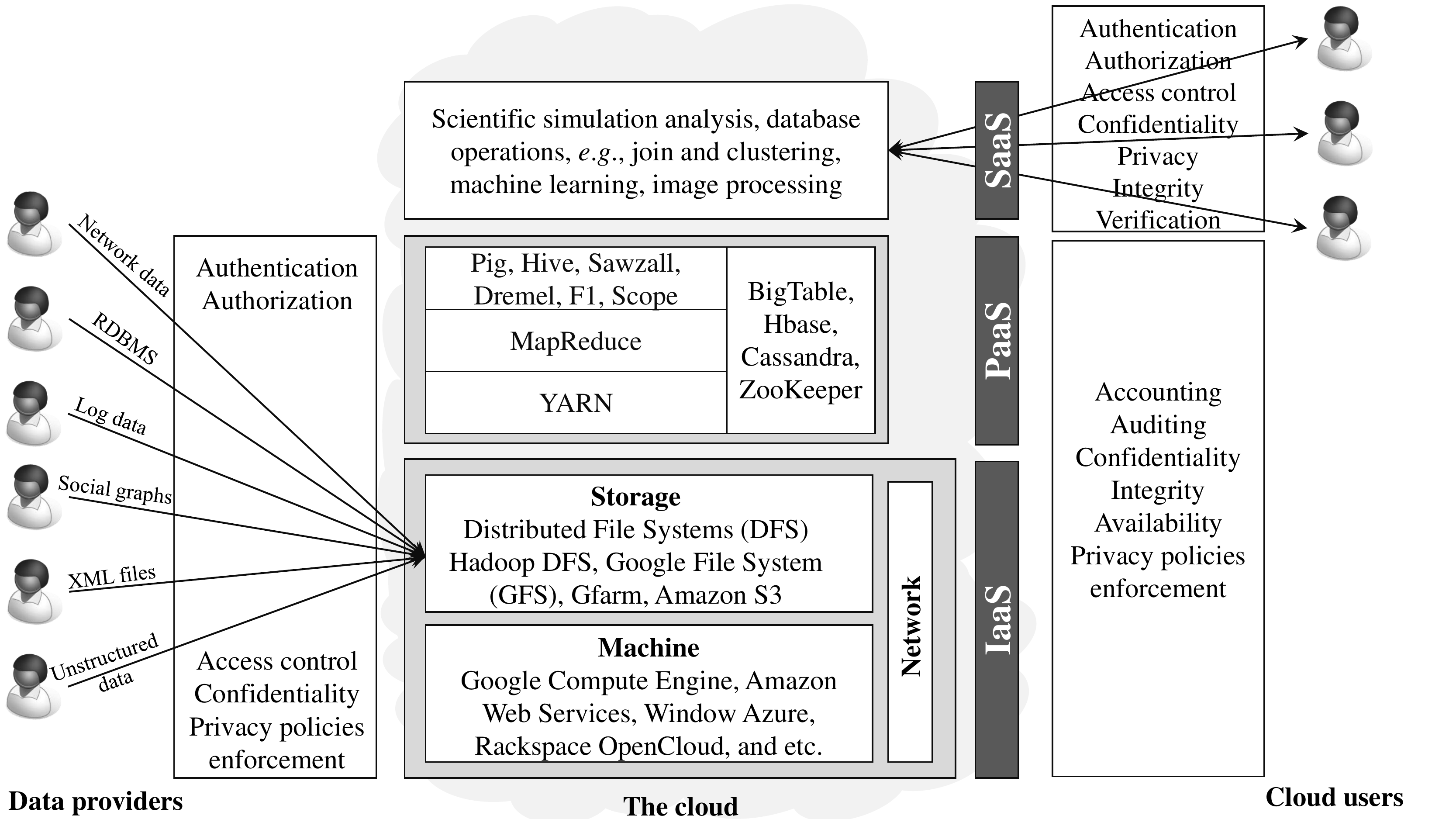}
\caption{Security requirements in MapReduce environment on the cloud. The figure shows a complete picture of the considered cloud structure, with different participating parties: data providers on the left, cloud provider in the middle and users on the right, and their specific security requirements. The figure also depicts various cloud levels and their relation to the security mechanisms.}
\label{fig:pic_mr+cloud}
\end{figure}

Dynamic deployment and configuration of mappers and reducers for each MapReduce computation specify distinct security requirements in MapReduce as compared to the classical parallel processing and cloud computing. Specifically, the allocation of different heterogeneous resources for different MapReduce computations, variability in the locations of resources, and different trust levels for different jobs are a few parameters that complicate the security of MapReduce. In order to overcome security challenges in MapReduce computations, a secure MapReduce must provide authenticated and authorized access, confidentiality (a secure storage and computation), integrity (a fair execution and storage), and accounting-auditing of data and computations. Next, we present the security requirements needed in a MapReduce computation. Most of the security requirements and corresponding cloud layers are depicted in Figure~\ref{fig:pic_mr+cloud}. As can be seen from Figure~\ref{fig:pic_mr+cloud}, we do not deal with issues of security above the MapReduce layer such as those of Pig, Hive, or big-data applications.

\medskip\noindent\textbf{Authentication, authorization, and access control of mappers and reducers.} Authentication provides a way to identify an adversarial mapper, reducer, or user. In other words, authentication is a process by which only those mappers and reducers that have rights to process data perform assigned tasks, and all the other mappers and reducers who are not allowed to access data and the framework are denied. Once mappers and reducers are authenticated, authorization of mappers and reducers allows them to access and process data by investigating their access privileges to that data. Access control provides pre-configured policies that restrict an unauthorized user to access data and to access the framework. An adversarial mapper, reducer, or user \emph{mimics} a legal mapper, reducer, or user, while breaching authentication and authorization. Attacks on authentication and authorization mechanisms are impersonation and replay attacks. The framework and data cannot be processed by any adversarial mapper, reducer, and user, when authentication, authorization, and access control of mappers and reducers are functioning properly.

\medskip\noindent\textbf{Availability of data, mappers, and reducers.} Data, mappers, and reducers should be available to authenticated and authorized users without delay. An adversarial code may make mappers-reducers and the network too busy so that they cannot process data and available to transfer data, respectively. An attack on availability of data, mappers, and reducers can \textit{interrupt} MapReduce computations. Specifically, an attack such as Denial-of-Service, is an attack on availability of data, mappers, and reducers. As an example, a single adversarial user in multi-users cloud-based MapReduce environment can considerably impact job completion times of the entire cluster even when using less than 10$\%$ of cluster resources~\cite{6906760}.

\medskip\noindent\textbf{Confidentiality of computations and data.} Confidentiality of computations and data refers to the protection of computations and data, which may be in-transit from the user's location to the location of computations or may be stored at public clouds, from unauthorized users and the public cloud itself. An attack on confidentiality of computations and data is \emph{interception} (\textit{i}.\textit{e}., eavesdropping and man-in-the-middle attacks). By ensuring confidentiality, one cannot intercept computations and data during the transmission and following transmission on public clouds themselves.

\medskip\noindent\textbf{Integrity of computations and data.} Integrity of a computation refers to a \emph{fair} transmission (of a MapReduce computation from the user's location to the locations of the computation) and execution of mappers and reducers. Similarly, integrity of data refers to a \emph{fair} transmission and storage of data. An attack on integrity of computations and data modifies computations or the data. To overcome such attacks, integrity should be preserved such that any modification and loss of computations and data by public clouds or an adversarial user is detected~\cite{DBLP:journals/fgcs/XiaoX14,DBLP:conf/acsac/WeiDYG09}. However, the amount of data hurdles to check the integrity of computations and the integrity of the whole data at a single computing node~\cite{DBLP:conf/acsac/WeiDYG09}.

\medskip\noindent\textbf{Verification of outputs.} Verification of outputs is a difficult task in MapReduce due to a massive parallel processing and a huge amount of input/output data. Verification ensures completeness (\textit{i}.\textit{e}., all the possible outputs are produced by mappers and reducers by processing of assigned input data without any tampering), correctness (\textit{i}.\textit{e}., all the outputs are legitimate and generated from assigned input data without any tampering), and freshness (\textit{i}.\textit{e}., all the outputs are new, not containing any old output, and generated from input data without any tampering) of outputs~\cite{DBLP:conf/codaspy/WeiYX13}.

\medskip\noindent\textbf{Accounting and auditing of computations and data.} Accounting of computations and data assists in locating mappers or reducers who are taking adversary actions over data with verifiable evidences~\cite{DBLP:journals/fgcs/XiaoX14}. Auditing of computations and data produces details of actions taken by mappers and reducers over data. Specifically, auditing investigates accounted data and provides verifiable evidences of what, when, initiated by whom, and how actions happened over which part of the data. An auditing process involves three parties, the first is data providers, the second is public clouds, and the third is the auditor who performs auditing. Hence, it is required to restrict data with fine-grained access controls, because an attacker who impersonates a legitimate auditor may understand the computation (better than the auditor) and reveal sensitive data.

\subsection{Adversarial Models for MapReduce Security}
\label{sbsec:Adversary Models for MapReduce Security}
There are numerous possible adversarial models in security, and thus, we will concentrate only on a limited set of adversarial models involved in MapReduce security.

\medskip\noindent\textbf{Honest-but-Curious adversary.} An honest-but-curious adversary executes a MapReduce job correctly and does not interfere with the job as well as data; however, it performs some extra computations for understanding the whole data and the job. For example, a public cloud does not interfere a MapReduce job and data, but it can observe the job and data. This type of adversary is also relevant to Privacy in MapReduce, see Section~\ref{subsection:Adversary Models for MapReduce Privacy}.

\medskip\noindent\textbf{Malicious adversary.} A malicious adversary can execute any computation for stealing, corrupting, and modifying data as well as the original MapReduce computation~\cite{2012sapsc,DBLP:dblp_conf/sose/DingWSFGZ13,CCGrid14loganalysis,DBLP:journals/fgcs/XiaoX14}. For example, a malicious mapper or reducer may not perform a computation or may provide wrong outputs (to users).

Malicious adversaries in distributed settings, like MapReduce, are divided into two types, as follows: (\textit{i}) \textit{non-collusive malicious adversary}: a non-collusive malicious adversary works independently, and hence, provides wrong outputs without consulting other malicious adversaries. In this case, if an identical task is assigned to two nodes and at least one of them is non-collusive, then the malicious behavior of the node can be easily detected by comparing their outputs; (\textit{ii}) \textit{collusive malicious adversary}: a collusive malicious adversary communicates with all the other malicious adversaries before providing outputs~\cite{DBLP:conf/IEEEcloud/WangW11,DBLP:conf/bigdataconf/WangWSDD13,DBLP:conf/acsac/WeiDYG09}. In this case, when a collusive (malicious) adversary is assigned a task, it consults other collusive adversaries to find if they are assigned an identical task. If yes, then all the collusive adversaries provide an identical wrong output, which make it harder to detect them.

Note that honest-but-curious and malicious adversaries are assumed to be polynomial-computationally-bounded adversaries, \textit{i}.\textit{e}., the adversaries cannot perform brute-force attack.

\medskip\noindent\textbf{Knowledgeable adversary.} A knowledgeable adversary is assumed to be knowledgeable with respect to the cloud structure, MapReduce algorithms, and implementation of mappers-reducers. In other words, a knowledgeable adversary, which may be the cloud provider or any user, is considered to be capable of leveraging any security (and privacy) threats in the framework.

\medskip\noindent\textbf{Network and nodes access adversary.} A network and nodes access adversary is assumed to have an access to network and nodes, though it does not have privileged accounts on the nodes. Privileged account on the nodes can be compared to the adversary owning the cloud, which is a situation where cloud user cannot be protected.

\subsection{Proposed Solutions for Securing MapReduce}
\label{subsec:Proposed Solutions for Security in MapReduce}
In this section, we provide a brief description of some existing security solutions for MapReduce. Reviewed security algorithms, protocols and frameworks are summarized in Table~\ref{table:Summary of security algorithms1}.

\subsubsection{Authentication, authorization, and access control based approaches}
\label{subsec:Access control based approaches}

An authentication mechanism for Hadoop using Kerberos and three special types of tokens, namely \textit{delegation token}, \textit{block access token}, and \textit{job token}, is presented~\cite{o2009hadoop,das2013adding}. The communication between a user and HDFS is divided into two parts: (\textit{i}) a user accesses NameNode using Hadoop's remote procedure call (RPC) libraries, and all RPCs connect using \emph{Simple Authentication and Security Layer} that uses Kerberos, DIGEST-MD5, or a delegation token and (\textit{ii}) a user accesses DataNodes using a streaming socket connection that is secured using a block access token. The working of the three types of tokens is as follows:

\begin{figure}[h]
    \begin{minipage}[t]{0.45\linewidth}
    \centering
    \includegraphics[scale=0.5]{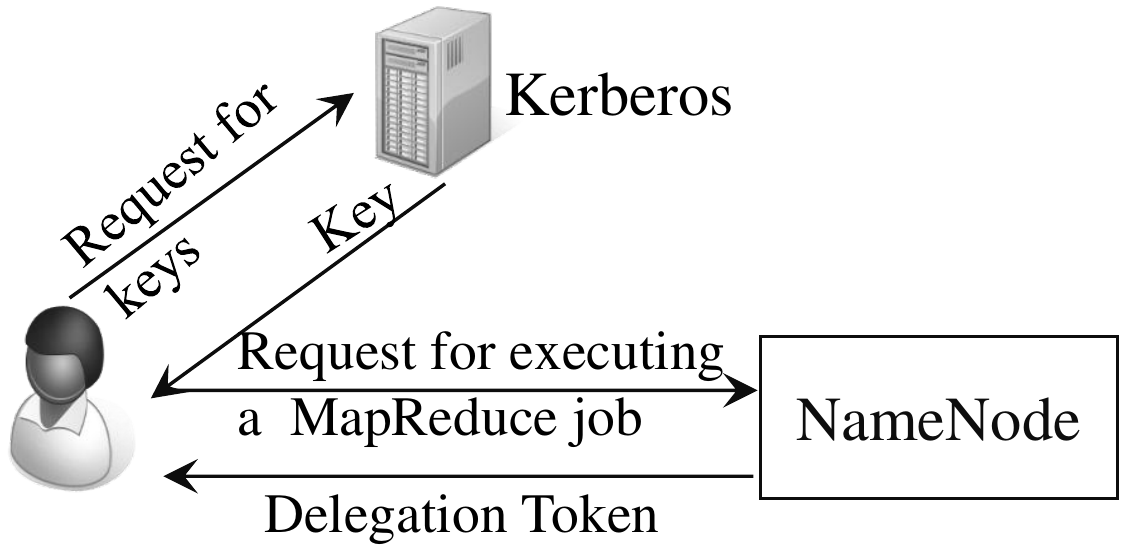}
    \subcaption{Delegation token.}
    \label{fig:Delegation token}
    \end{minipage}
\quad\quad
    \begin{minipage}[t]{0.49\linewidth}
    \centering
    \includegraphics[scale=0.5]{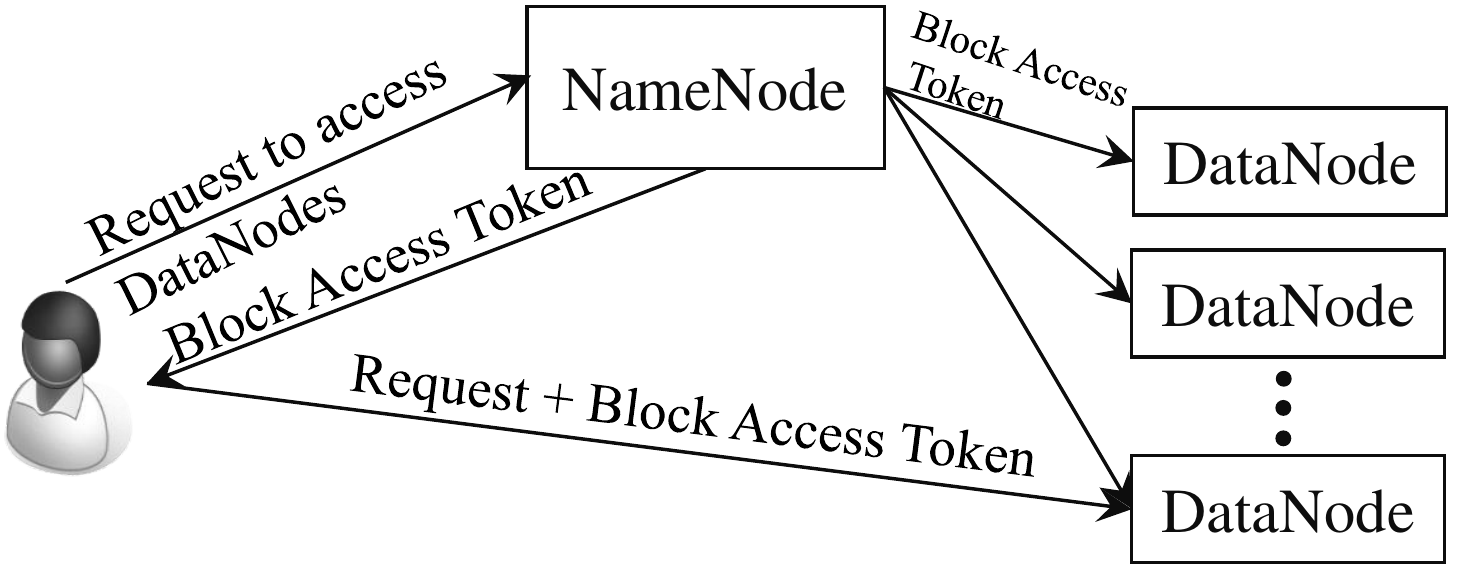}
    \subcaption{Block access token.}
    \label{fig:Block access token}
    \end{minipage}
\caption{Authentication to Hadoop using Kerberos and two types of tokens.}
\end{figure}

\noindent \textit{Delegation token.} A delegation token is a secret key between a user and NameNode. After authenticating a user, a delegation token is generated by NameNode using Kerberos, and the token is used for subsequent authentication of the user by NameNode without involving Kerberos; see Figure~\ref{fig:Delegation token}.

\noindent \textit{Block access token.} A block access token provides authentication policies to DataNodes by passing authorization information from NameNode to DataNodes and is generated by NameNode using a symmetric-key scheme where NameNode and all of the DataNodes share a secret key; see Figure~\ref{fig:Block access token}. When a user wants to access a file in HDFS, it requests NameNode for block ids and locations of the file. In response, NameNode sends block ids and the locations with a block access token for each block, if NameNode verifies authenticity and authority of the user. The user sends the block id with the block access token to the corresponding DataNode, and the DataNode verifies the block access token before allowing access to that block.

\noindent \textit{Job token.} A job token is generated by JobTracker in the form of a secret key, when a MapReduce job is submitted. JobTraker stores the job token for each job as a part of the job and distributes it to TaskTrackers. The job token is used to authenticate tasks at TaskTrakers.

MapReduce paradigm itself is security agnostic, and this lead to an initial version of Hadoop implementation to have no built-in security mechanism. However, as MapReduce in general and, specifically, Hadoop gained wide-spread usage, the need for security has become more and more acute. This led to a number of recent projects, which we will review next, trying to add different types of security aspects to Hadoop.

\medskip\noindent\textbf{Apache Knox.} Apache Knox~\cite{knox} is a stateless reverse proxy framework that provides gateway-level security and a single access point to a single Hadoop cluster or multiple Hadoop clusters. Apache Knox provides a monitoring of the system, authentication, federation of authenticated users, authorization, and auditing. It has several advantages, such as: integration with enterprise identity management solutions, it hides details of Hadoop cluster deployment, simplifies the number of services that clients need to interact with, limits numbers of access point to Hadoop clusters, scales linearly by adding more Knox nodes as the load increases.

\medskip\noindent\textbf{Apache Sentry.} Apache Sentry~\cite{sentry} is a system for fine-grained, multi-tenant administration, and role-based authorization of an access to data and metadata in a Hadoop cluster. Apache Sentry can be integrated in relational data model for Apache Hive, Cloudera Impala, and hierarchical data model used by Apache Solr. Sentry allows access control at the server, database, table, and view scopes. It also allows different privileges for \texttt{select}, \texttt{insert}, \texttt{create}, and \texttt{modify}. Sentry defines policies for accessing resources. On receiving a request from a user, Hive/Impala/Solr asks Sentry for validating the request. Sentry builds a map of privileges allowed for the requesting user and then determines whether the given request should be allowed, and then allows or prohibits the user access based on decisions by Sentry.

\medskip\noindent\textbf{Apache Ranger.} Apache Ranger~\cite{ranger} provides a centralized and comprehensive platform for securing Hadoop. Specially, Apache Ranger provides: (\textit{i}) authentication: by Kerberos and secured by Apache Knox, (\textit{ii}) authorization, (\textit{iii}) fine-grained access control: by role-based access control, attribute-based access control, etc., (\textit{iv}) auditing of HDFS, Hive, and HBase, and (\textit{v}) data protection: by wire encryption, volume encryption, and file/column encryption.

\medskip\noindent\textbf{Project Rhino.} Project Rhino~\cite{rhino}, an initiative by Cloudera and Intel, is an open source for enhancing existing data protection in the Hadoop \textit{stack}. Specifically, Project Rhino provides framework support for encryption, key management, and a common authentication-authorization module with single sign on. Recently, Project Rhino added cell level encryption and fine-grained access control to HBase 0.98, and encryption to data at-rest (data stored on persistent storage) in Apache Hadoop. Note that data encryption in Hadoop requires encryption of data at-rest and in-transit; however, except Project Rhino, most Hadoop components provide encryption for data in-transit only.

\medskip\noindent\textbf{Apache Accumulo.} Apache Accumulo~\cite{accumulo} is not a security framework like Apache Knox, Apache Sentry, Apache Ranger, and Project Rhino. However, Apache Accumulo improves Google BigTable~\cite{BigTable} design by introducing \textit{cell-based access control}, which emphasizes us to mention Apache Accumulo in this paper. Apache Accumulo stores a logical combination of security labels that must be satisfied at query time in order for keys and values to be returned as part of a user request. This allows users to see only those keys and values for which they are authorized. Apache Accumulo is built on top of Apache Hadoop, Zookeeper, and Thrift.

\bgroup
\def\arraystretch{1.105}
\scriptsize
\centering
\begin{longtable}[t]{|p{3.3cm}|l|l|l|l|l|l|l|l|l|l|l|l|l|l|l|l|l|}
\caption{Summary of security algorithms, protocols, and frameworks for MapReduce.}
\label{table:Summary of security algorithms1}\\\hline

%


   \multirow{2}{3.3cm}{Algorithms/Protocols/\\Frameworks} & \multirow{2}{*}{\rotatebox{90}{\parbox{2.25cm}{Authentication,\\ authorization}}} & \multirow{2}{*}{\rotatebox{90}{\parbox{2.25cm}{Access Control}}} & \multicolumn{2}{ c| }{Confidentiality} & \multicolumn{2}{ c| }{Integrity} & \multirow{2}{*}{\rotatebox{90}{\parbox{2.25cm}{Availability}}} & \multirow{2}{*}{\rotatebox{90}{\parbox{2.25cm}{Accounting}}} & \multirow{2}{*}{\rotatebox{90}{\parbox{2.25cm}{Auditing}}} & \multirow{2}{*}{\rotatebox{90}{\parbox{2.25cm}{Cloud structure}}} & \multicolumn{6}{ c| }{Attack handling} \\

 \hhline{~~~|-|-|-|-|~~~~|-|-|-|-|-|-|-|}  & {~} & {~}  & {\rotatebox{90}{\parbox{2cm}{Data}}} & {\rotatebox{90}{\parbox{2cm}{Computation}}} & {\rotatebox{90}{\parbox{2cm}{Data}}} & {\rotatebox{90}{\parbox{2cm}{Computation}}} & & & & & {\rotatebox{90}{\parbox{2cm}{Impersonation}}} & {\rotatebox{90}{\parbox{2cm}{DoS}}} & {\rotatebox{90}{\parbox{2cm}{Replay}}} & {\rotatebox{90}{\parbox{1cm}{Eavesdropping}}} & {\rotatebox{90}{\parbox{2cm}{MiM}}} & {\rotatebox{90}{\parbox{2cm}{Repudiation}}} \\ \hline

   ~\cite{o2009hadoop},~\cite{das2013adding} & \checkmark & \checkmark & & & & & & & & S\footnote{S: Single cloud.} &\checkmark & & & & & \\\hline

   Apache Knox~\cite{knox} & \checkmark  & \checkmark  &   &  & & & & & \checkmark & M\footnote{M: Multiple clouds.} & & & & & & \\\hline

   Apache Sentry~\cite{ranger} & \checkmark  & \checkmark  &   &  & & & & &  & S & & & & & & \\\hline

   Apache Ranger~\cite{ranger} & \checkmark  & \checkmark  & \checkmark  &  & & & & & \checkmark & M & & & & & & \\\hline

   Project Rhino~\cite{rhino} & \checkmark  & \checkmark  & \checkmark  &  & & & & & \checkmark & S & \checkmark & & & & & \\\hline

   Apache Accumulo~\cite{accumulo} &   & \checkmark  &  &  & & & & &  & S &  & & & & & \\\hline

   Airavat~\cite{DBLP:conf/nsdi/RoySKSW10} & \checkmark  & \checkmark  & \checkmark  &  & & & & & & S &\checkmark & & & & & \\\hline

   Khaled et al.~\cite{DBLP:conf/cloudcom/KhaledHKHT10} &  & \checkmark & & & & & & & & & &  & & & & \\\hline

   Vigiles~\cite{2014vigiles} & \checkmark & \checkmark & & & & & & & & & & & & & &  \\\hline

   GuardMR~\cite{DBLP:conf/ccs/UlusoyCFKP15} & \checkmark & \checkmark & & & & & & & & & & & & & &  \\\hline

   G-Hadoop~\cite{DBLP:journals/jcss/ZhaoWTCSRKSG14} & \checkmark & \checkmark & & & & & & & & &\checkmark & & \checkmark & & \checkmark &  \\\hline

   SecDM~\cite{DBLP:conf/dasc/ShenZYYWZ11} & \checkmark & \checkmark & \checkmark & \checkmark & \checkmark & \checkmark & &  & & M & & & &\checkmark &\checkmark & \\\hline

   iBigTable~\cite{DBLP:conf/codaspy/WeiYX13} & & & & & \checkmark & \checkmark& & & &  & \checkmark& & & & & \\\hline

   Lin et al.~\cite{DBLP:conf/aina/LinSTL12} & & \checkmark & & & & & & & &S & & & & \checkmark\footnote{Partially.} &\checkmark\footnote{Partially.} & \\\hline

   SAPSC~\cite{2012sapsc} &\checkmark & & & & & & & & & M & & & &\checkmark & & \\\hline

   ClusterBFT ~\cite{13middleware} & & & & & \checkmark  & &  & & & S, H\footnote{H: Hybrid cloud.} & \checkmark & & & & & \\\hline

   Moca et al.~\cite{11desktopgridmrsecurity} & & & & &\checkmark &\checkmark & & & & & & & & & &\checkmark  \\\hline

   SecureMR~\cite{DBLP:conf/acsac/WeiDYG09} & & & \checkmark& &\checkmark &\checkmark & \checkmark& & & S & \checkmark&\checkmark & \checkmark& \checkmark& & \checkmark\\\hline

   AccountableMR ~\cite{DBLP:journals/fgcs/XiaoX14} & & & & & \checkmark & \checkmark & &\checkmark & & S & & & & & & \\\hline

   VIAF~\cite{DBLP:conf/IEEEcloud/WangW11} & & & & &\checkmark & \checkmark& & & & S & \checkmark & & &  & & \\\hline

   CCMR~\cite{ResultIntegritysinglepublic} & & & & &\checkmark & \checkmark& & & & M & \checkmark & & &  & & \\\hline

   IntegrityMR ~\cite{DBLP:conf/bigdataconf/WangWSDD13} & & & & &\checkmark & \checkmark& & & & M & \checkmark & & &  & & \\\hline

   VAWS~\cite{DBLP:journals/ieicet/DingWWCFX14} & & & & & \checkmark  & \checkmark & & & & S & \checkmark  & & & & & \\\hline

   Hatman~\cite{DBLP:conf/IEEEcloud/KhanH12} & & & & &\checkmark & & & & & S &  & & &  & \checkmark\footnote{By assumptions of the algorithms.} & \\\hline


TrustMR~\cite{DBLP:conf/bigdataconf/UlusoyKP15} & & & & & & \checkmark& & & & S &  & & &  & & \\\hline

   TS-TRV~\cite{DBLP:dblp_conf/sose/DingWSFGZ13} & & & & &\checkmark & \checkmark& & & & S & \checkmark & & &  & & \\\hline

   Log-based~\cite{CCGrid14loganalysis} & \checkmark & & & &\checkmark & \checkmark& & \checkmark & \checkmark & S & \checkmark & & &  & & \\\hline

   Watermarking based~\cite{DBLP:conf/ccgrid/HuangZW12,DBLP:conf/sose/DingWCTFS14} & & & & &\checkmark & \checkmark& & & & S & \checkmark & & &  & & \\\hline

Accountable MapReduce (RBAC)~\cite{7363786} & \checkmark & \checkmark & \checkmark & & & & & \checkmark & \checkmark & S & & & & & & \\\hline

\end{longtable}
\egroup


\medskip\noindent\textbf{Airavat.} Airavat~\cite{DBLP:conf/nsdi/RoySKSW10} is a system that provides mandatory access control together with differential privacy for data protection. Airavat is the first system that provides a complete solution for data privacy and secure computations in MapReduce environments. To ensure that untrusted mappers do not leak data outside the cluster, Airavat uses mandatory control system to disallow direct access of mappers to the data and to the network. The system is based on Security-Enhanced Linux (SELinux)\footnote{http://selinuxproject.org/} and requires a unified deployment of the secure Linux system on all the nodes of the cluster.

\begin{figure}[h]
\centering
\includegraphics[scale=0.4]{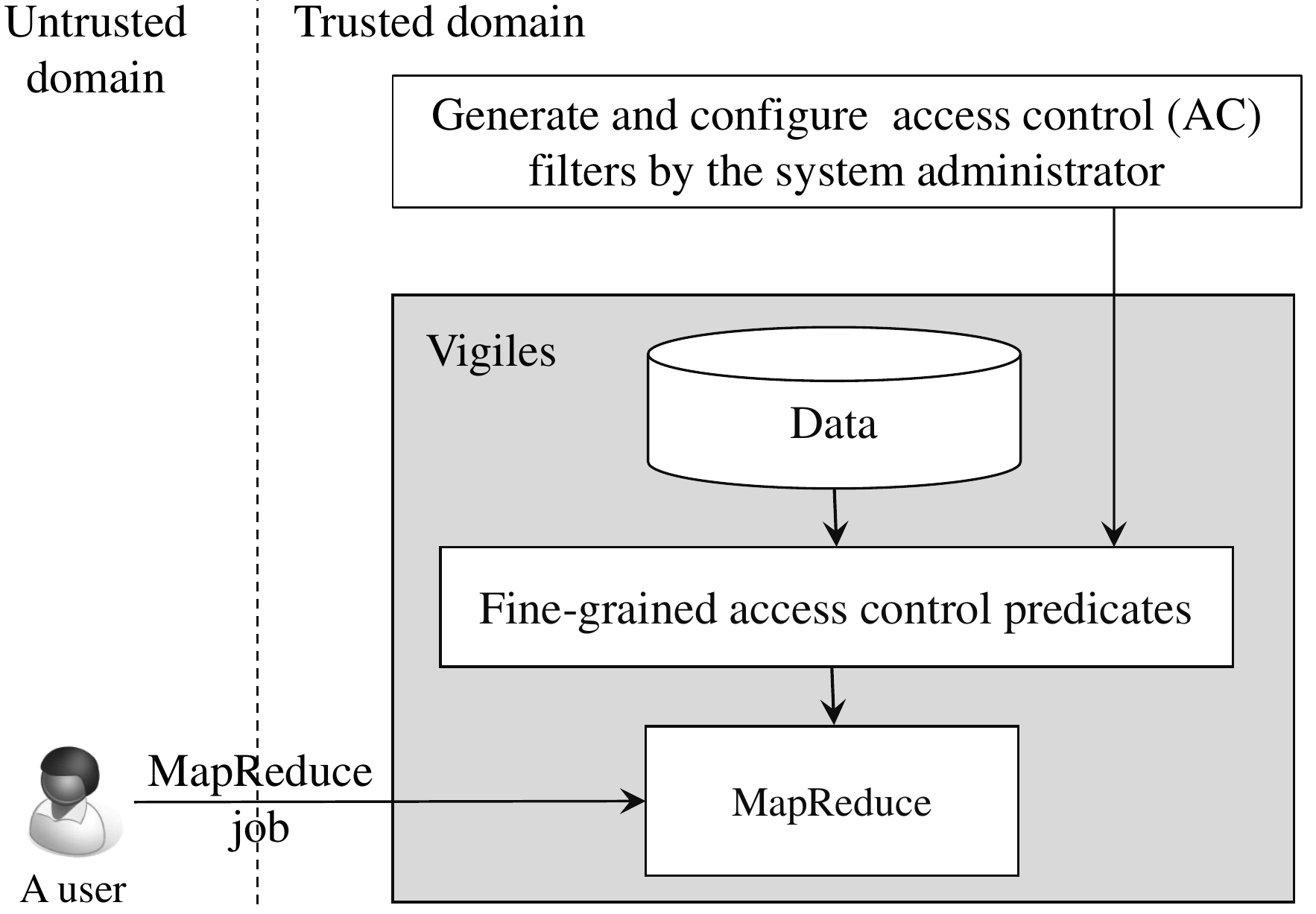}
\caption{Vigiles access control mechanism.}
\label{fig:vigiles}
\end{figure}

\medskip\noindent\textbf{Vigiles.} Vigiles~\cite{2014vigiles} provides a fine-grained access control mechanism for read operations in MapReduce without modifying computations. Vigiles works as a middleware architecture that stays between untrusted users and MapReduce environment; see Figure~\ref{fig:vigiles}. Since Vigiles supports only read operation, it is required to remove sensitive data from outputs (of the read operation). A system administrator manages HDFS and creates access control filters, which provide only authorized data to users, based on the given configurations. Access control filters are used to remove sensitive data by following a 3-phase procedure: decompose, fetch, and action. When users provide MapReduce computations, Vigiles executes them by following access control policies and remove sensitive data from the outputs. On the negative side, ad-hoc data types, append and delete operations are not allowed in Vigiles. Nevertheless, it does not require any modification of a MapReduce computation.

\medskip\noindent\textbf{GuardMR.} GuardMR~\cite{DBLP:conf/ccs/UlusoyCFKP15} allows ad-hoc data types and provides access to the record dynamically. GuardMR is composed of two main components: (\textit{i}) access control module performs the administrative functions that allow to add new data types and preprocessing functions after a proper security analysis; and (\textit{ii}) reference monitor enforces specified security policies to the underlying MapReduce system after consulting the access control module, resulting in an authorized view of data.

\medskip In~\cite{DBLP:conf/cloudcom/KhaledHKHT10}, an access control and enforcement policy based architecture for Resource Description Framework (RDF)~\cite{RDF,kaoudi2014rdf} is proposed, where the system administrator generates an access token for securely accessing data based on the request of users. The token prevents access to the entire data. Six types of secure data accesses are suggested: predicate data access, subject and object data access with or without predicates, and subject model level access. Since the format of RDF is not suitable for a MapReduce computation, a two-layered system is also proposed, where the first layer (data processing layer) converts RDF to $N$-Triple format~\cite{RDF-N-triple}, and the second layer (query processing layer) is responsible for executing a MapReduce computation. The query processing layer provides outputs regarding an access token. The query processing layer first rewrites a query that satisfies an access token, then performs a MapReduce job according to the rewritten query, and finally performs one or more additional MapReduce jobs to remove sensitive data from outputs according to the access token.

Authentication of malicious users based on storing communication between user and NameNode, and between user and DataNodes is suggested in~\cite{CCGrid14loganalysis}. The approach stores IP addresses, port numbers, and socket connection related system calls. In~\cite{DBLP:journals/jcss/ZhaoWTCSRKSG14}, a security model for geographically distributed Hadoop, called G-Hadoop~\cite{DBLP:journals/fgcs/WangTRMSCC13}, is presented. The model uses two types of tokens, namely proxy token and slave token for the purpose of authentication. A proxy token (contains its expiration time, identity of certificate authority (CA) server, the public key of the master process, and a random message generated by the CA server) is used by slave nodes for authenticating the master process. A slave token (contains identity of the CA server and the public key of the corresponding slave node) is used by the master process for authenticating a slave node. In~\cite{DBLP:conf/isi/UlusoyKTK15}, the authors suggested a honey-pot-based mechanism for detecting an unauthorized data access. Honey data is deliberately produced and mixed in the original data. However, an authorized user never accesses the honey-pot data during a MapReduce job. Since attackers access all the parts of data, it leads to an alarm with a high probability.

\subsubsection{An encryption-decryption based approach for data transmission}
\label{subsec:Encryption-decryption based approaches}
A secure data transmission and security storage of the data (reviewed in Section~\ref{section:Privacy Aspects in MapReduce}) are critical issues when mappers and reducers that reside on two different clouds share data.

\medskip\noindent\textbf{SecDM.} Secure Data Migration (SecDM)~\cite{DBLP:conf/dasc/ShenZYYWZ11} provides a secure way for data transmission among mappers-reducers at two different clouds. The two master processes, which are located at two different clouds, create a Secure Socket Layer (SSL) connection between them, then send message authentication code with a timestamp and negotiate a random key. After authentication of two master processes is completed, mappers or reducers receive the locations of data in the other cloud. Data transmission (among mappers-reducers at two clouds) is carried out using the negotiated key, where encrypted data is transmitted with a hash value of data and a message authentication code, which prevents tampering.

\subsubsection{Approaches for security and integrity of storage}
\label{subsec:Approaches for security of storage}
\begin{figure}[h]
            \begin{minipage}[t]{0.45\linewidth}
            \centering
            \includegraphics[scale=0.4]{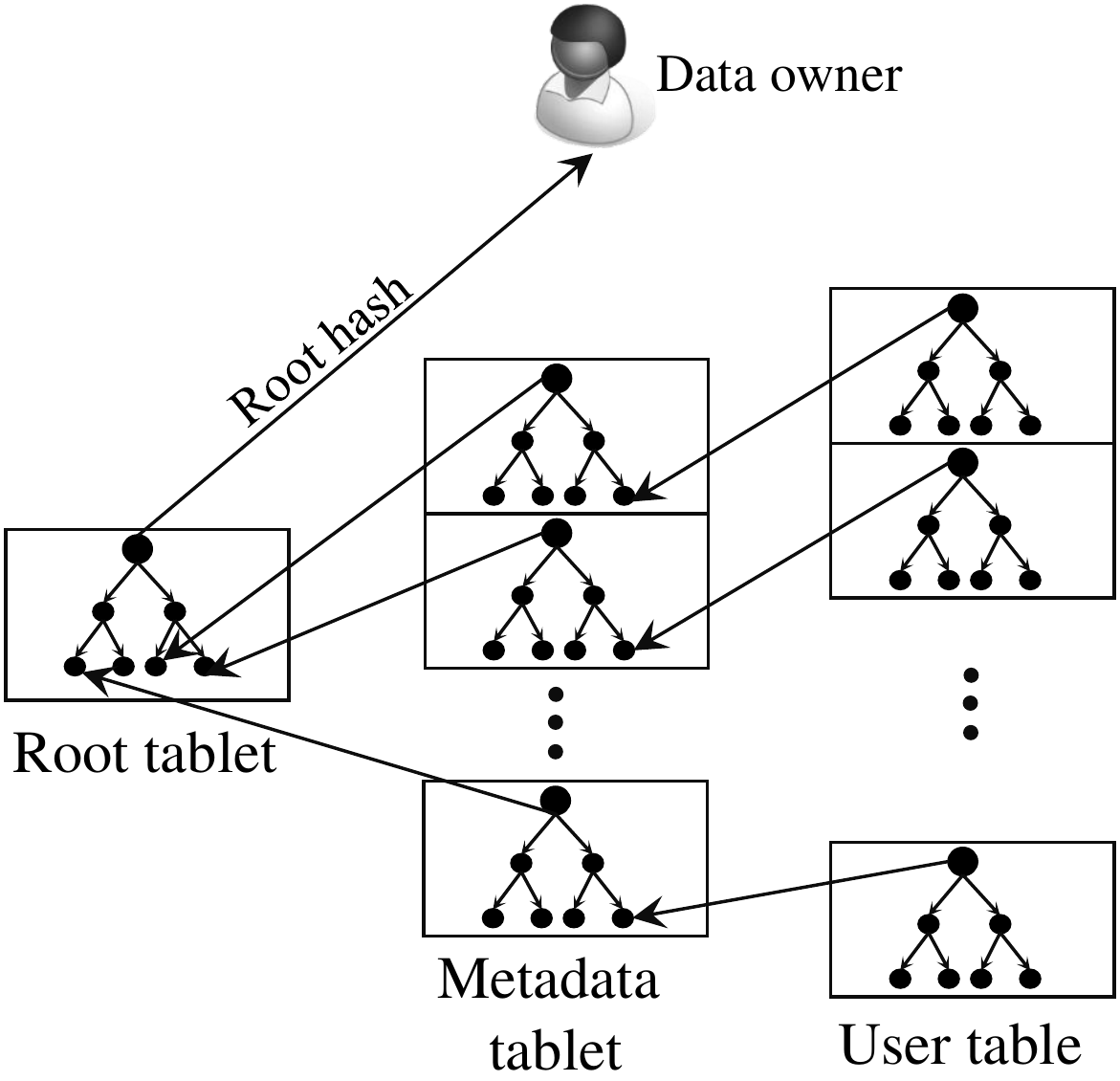}
            \subcaption{The first (centralized) approach.}
            \label{fig:ibigtable1}
            \end{minipage}
  \quad\quad
            \begin{minipage}[t]{0.49\linewidth}
            \centering
            \includegraphics[scale=0.4]{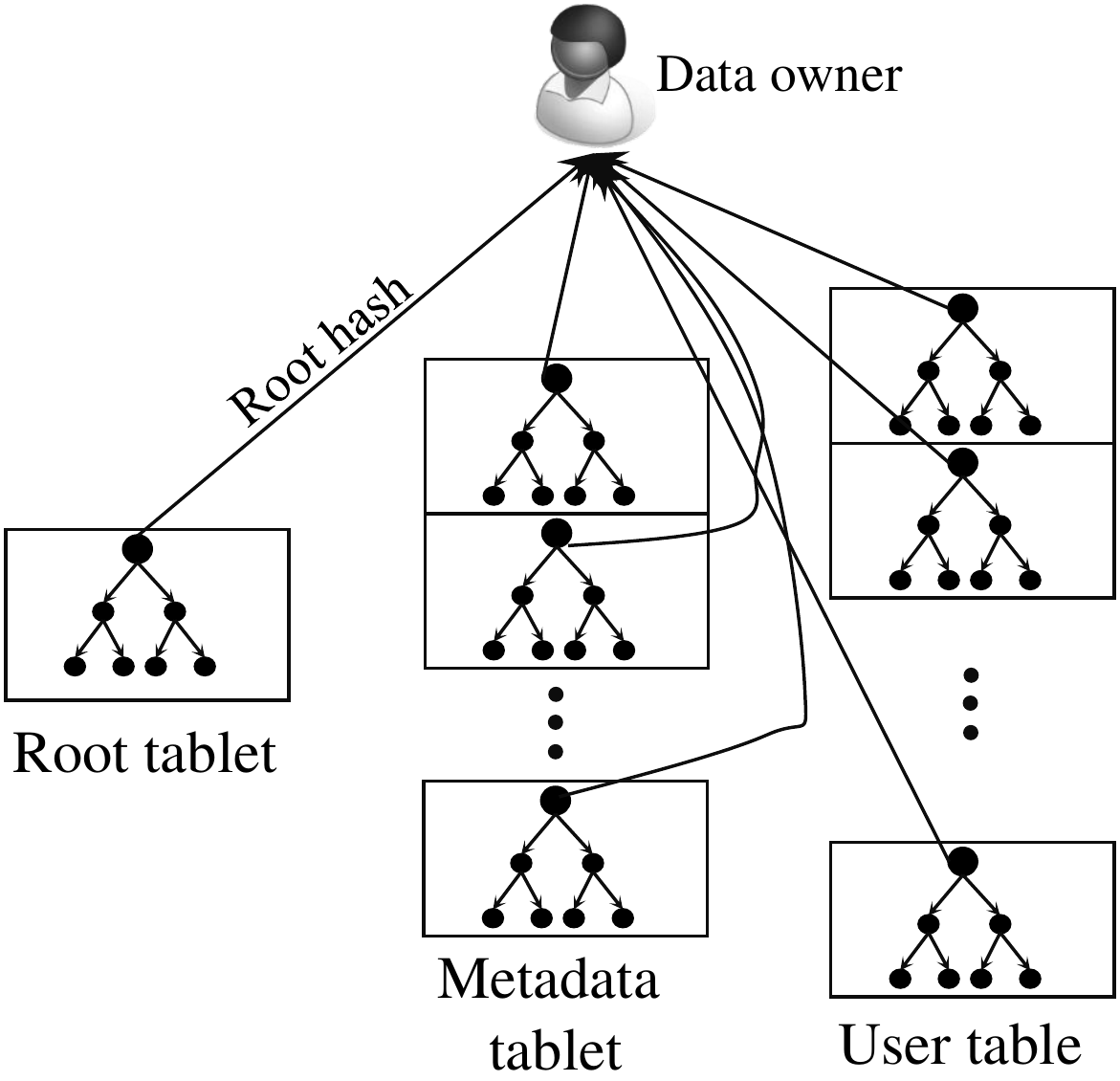}
            \subcaption{The second (distributed) approach.}
            \label{fig:ibigtable2}
            \end{minipage}
            \caption{Merkle Hash Tree based authenticated storage of data structures in iBigTable.}
\end{figure}

\medskip\noindent\textbf{iBigTable.} An enhancement of BigTable, called iBigTable~\cite{DBLP:conf/codaspy/WeiYX13}, ensures the integrity of data using decentralized authenticated data structures. Two approaches, see Figures~\ref{fig:ibigtable1} and~\ref{fig:ibigtable2}, are suggested for storing authenticated data that is used to verify the integrity of data. Both the approaches build a Merkle Hash Tree (MHT)~\cite{DBLP:conf/crypto/Merkle87} based authenticated data structure for the root tablet, the metadata tablets, and the user tables. MHT allows efficient and secure verification of contents of large-scale data, where non-leaf nodes are labelled with the hash of the labels of their child nodes.

In the first (centralized) approach, each user table and the metadata tablet stores its root hash of the authenticated data structure at the metadata tablets and the root tablet, respectively; see Figure~\ref{fig:ibigtable1}. The root hash of the root tablet is stored at the user-end. Whenever any data is updated at a user table, the corresponding authenticated data structure is also updated. An update of the authenticated data structure at a user table requires updates of the corresponding authenticated data structures at the metadata tablet, the root tablet, and the user-end. Such updates of authenticated data structures decrease the performance of BigTable and require the involvement of the user, the metadata tablets, and the root tablet. In the second (distributed) approach, a user stores the root hash of each tablet and each user table, see Figure~\ref{fig:ibigtable2}; and storing the root hash at the user-end increases the performance of iBigTable. In addition, it is not required to store the root hash of the authenticated data structures at the higher level.

In order to verify the integrity of data, three rounds of communication between a client and the servers are required. The first round requires communication between the user and the root tablet; the second round requires communication between the user and the metadata tablet; and the third round requires communication between the user and the user table. In each round, a tablet server generates and sends a \emph{verification object}, VO, which contains a set of hashes, for the data sent to the user. On receiving data and the corresponding VO, the user verifies the integrity of the received data. VOs from the root tablet and the metadata tablet allow accessing the metadata tablet and the user table, respectively.

\begin{figure}[h]
\centering
\includegraphics[scale=0.4]{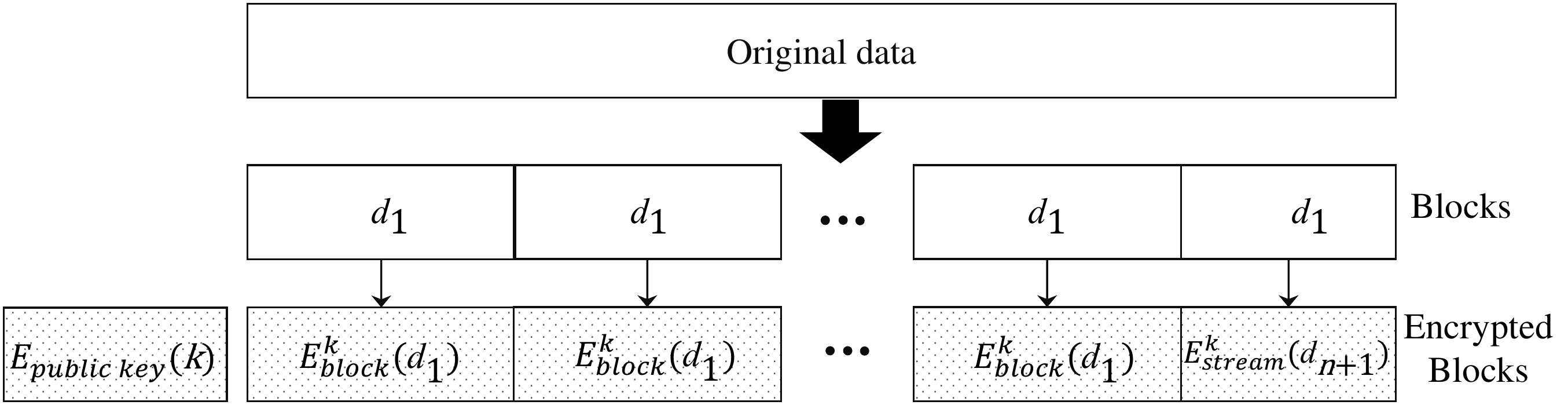}
\caption{A hybrid encryption scheme used to store data in HDFS.}
\label{fig:hadoop-hdfs-pic}
\end{figure}

\medskip\noindent\textbf{HDFS-RSA and HDFS-Pairing.} In order to store data in a confidential manner in HDFS, two approaches based on hybrid encryption are suggested in~\cite{DBLP:conf/aina/LinSTL12}. The hybrid encryption approaches, see Figure~\ref{fig:hadoop-hdfs-pic}, use a block cipher and a stream cipher (see Chapter 3 of ~\cite{william2006cryptography} for block ciphers and stream ciphers). Data is divided into fixed-sized blocks, $d_1, d_2, \ldots, d_n$, according to a block cipher, and the remaining part of data becomes block $d_{n+1}$. The blocks $d_1, d_2, \ldots, d_n$ are encrypted using a random key $k$ and a block cipher, and the block $d_{n+1}$ is encrypted using the key $k$ and a stream cipher. The key, $k$, is, then, encrypted using a public key scheme. The first approach, called \emph{HDFS-RSA}, uses the RSA encryption scheme and AES, and the second approach, called \emph{HDFS-pairing}, uses a pairing-based encryption scheme and AES. However, both the approaches are suitable for applications with a few write and many read operations.

\begin{figure}[h]
\centering
\includegraphics[scale=0.4]{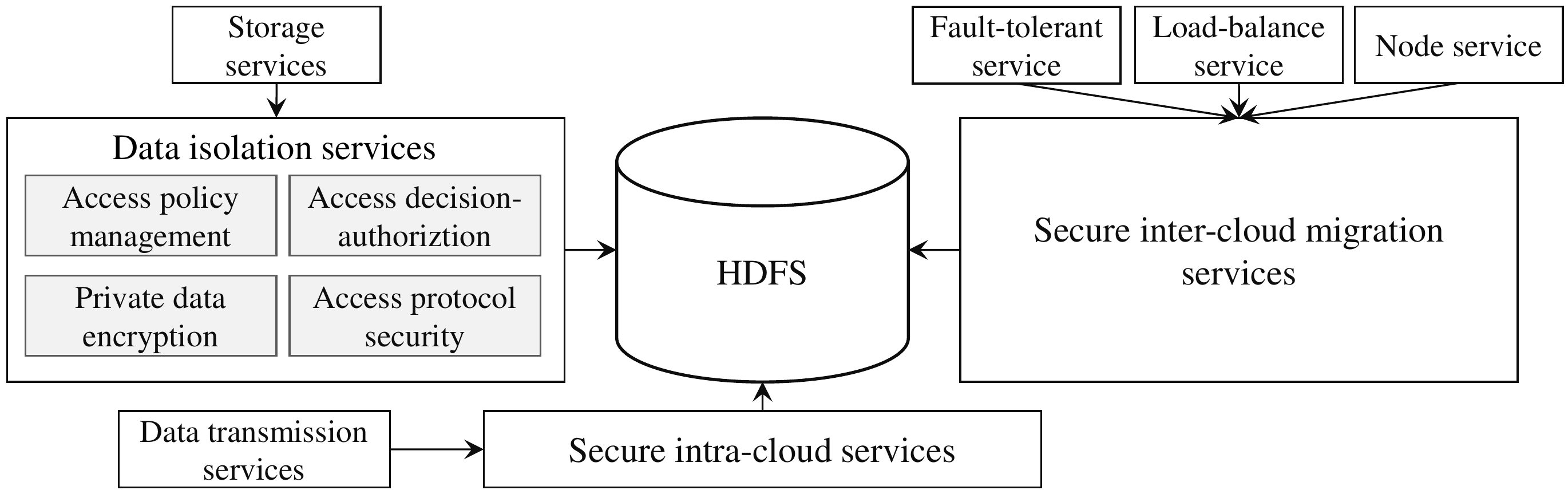}
\caption{SAPSC architecture for data security.}
\label{fig:sapsc}
\end{figure}

\medskip\noindent\textbf{SAPSC.} Security Architecture of Private Storage Cloud based on HDFS (SAPSC)~\cite{2012sapsc} provides an architecture for ensuring security of data stored in HDFS by \textit{data isolation service}, \textit{secure intra-cloud data migration service}, and \textit{secure inter-cloud migration service}; see Figure~\ref{fig:sapsc}. These three services are dependent on five major services of a distributed files system, as follows: (\textit{i}) fault-tolerant service, (\textit{ii}) storage service, (\textit{iii}) configuration of the system and management of the nodes, called node services, (\textit{iv}) data transmission service, and (\textit{v}) load balance service.

Data isolation services are invoked when a user reads/writes a file and are responsible for secure storage of data in HDFS and secure operations on data. Data isolation service is based on access policy management, access decision-authorization, access protocol security, and private data encryption. Secure intra-cloud data migration services do a secure data replication for the fault-tolerant service and transfer of data involved in a job, due to the load balance service or the node service. Secure inter-cloud data migration services do a secure data migration among clouds through the transmission service. In addition, three security policies are defined, as follows: (\textit{i}) a flexible access control policy based on role-based access control, (\textit{ii}) a label-based intra-cloud data replicating and restructuring policy, and (\textit{ii}) a temporary-ticket based parallel inter-cloud data transmission policy.

\subsubsection{Approaches for result verification and accounting}
\label{subsec:Approaches for result verification and accounting}
Several approaches for result verification are proposed based on redundancy of data and computations, trust management, and log analysis, which will be presented in this section.

\Subsubsubsection{Redundancy based approach}

A redundancy based approach replicates all or some of the tasks to multiple nodes and checks their outputs to find inconsistencies. Several approaches based on redundancy of tasks are reviewed below.

\medskip\noindent\textbf{ClusterBFT.} ClusterBFT~\cite{13middleware} uses the Byzantine Failure Tolerant (BFT)~\cite{DBLP:journals/toplas/LamportSP82} replication technique to cope with a situation where the cloud is trusted but there are potentially malicious nodes or users in a cluster. BFT replication is used for computational results verification and for overcoming untrusted, possibly malicious nodes. BFT replication techniques perform calculations in parallel on multiple replicas, then compare all the produced outputs to identify erratic behavioral nodes and decide a correct output based on a majority vote. However, current BFT replication techniques were developed for stand-alone servers and do not suit cloud-based computations, where data flow among different nodes and a computation consists of a number of stages to be performed on different nodes, as it is done in MapReduce. In order to overcome this gap, ClusterBFT algorithm adopts BFT replication for highly-scalable, distributed and high-granularity cloud computations. The algorithm identifies an optimal, according to heuristic function, subgraph of the computational flow that is verified by multiple replicas. The rest of the nodes participating in data flow are not replicated to avoid multiplication of verification messages and performance overhead. To reduce the volume of replicated data in verification phase, the algorithm uses digital digest of the data. In addition, the algorithm allows fault isolation and identification, \textit{i}.\textit{e}., identification of components that continuously return incorrect outputs and removing them from the job scheduling.

\begin{figure}[h]
           \begin{minipage}[t]{0.9\linewidth}
                \centering
                  \includegraphics[scale=0.4]{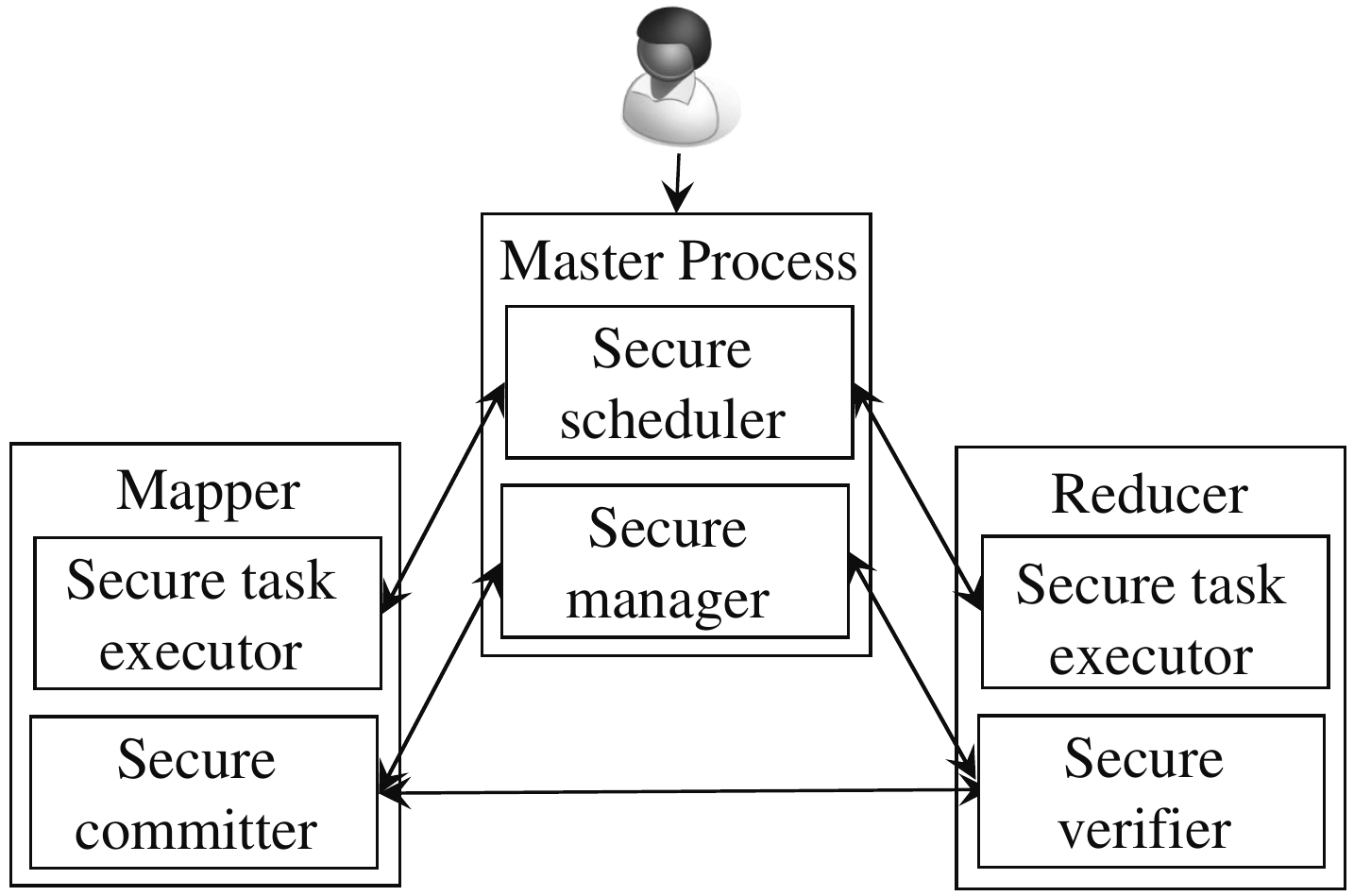}
              \subcaption{The framework.}
               \label{fig:secure-mr-pic1}
            \end{minipage}
            \quad
            \begin{minipage}[t]{0.49\linewidth}
                \centering
                   \includegraphics[scale=0.4]{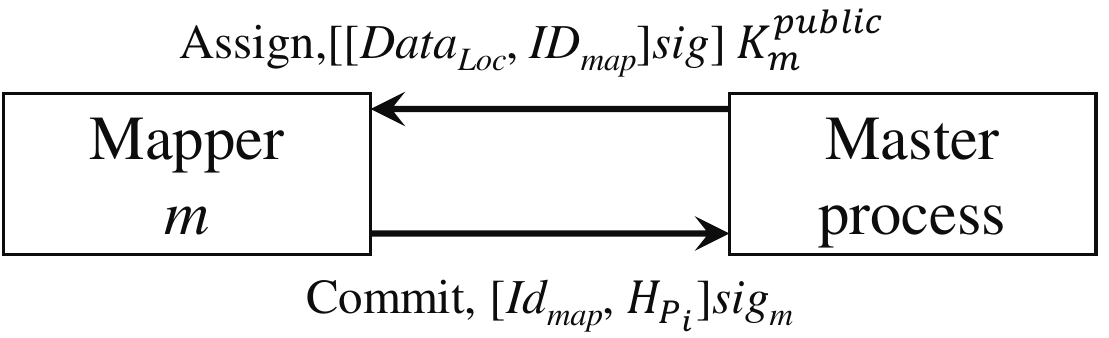}
                    \subcaption{Communication between the master process and a mapper, the commitment protocol.}
                \label{fig:secure-mr-pic2}
            \end{minipage}
            \quad
            \begin{minipage}[t]{0.49\linewidth}
                \centering
                   \includegraphics[scale=0.4]{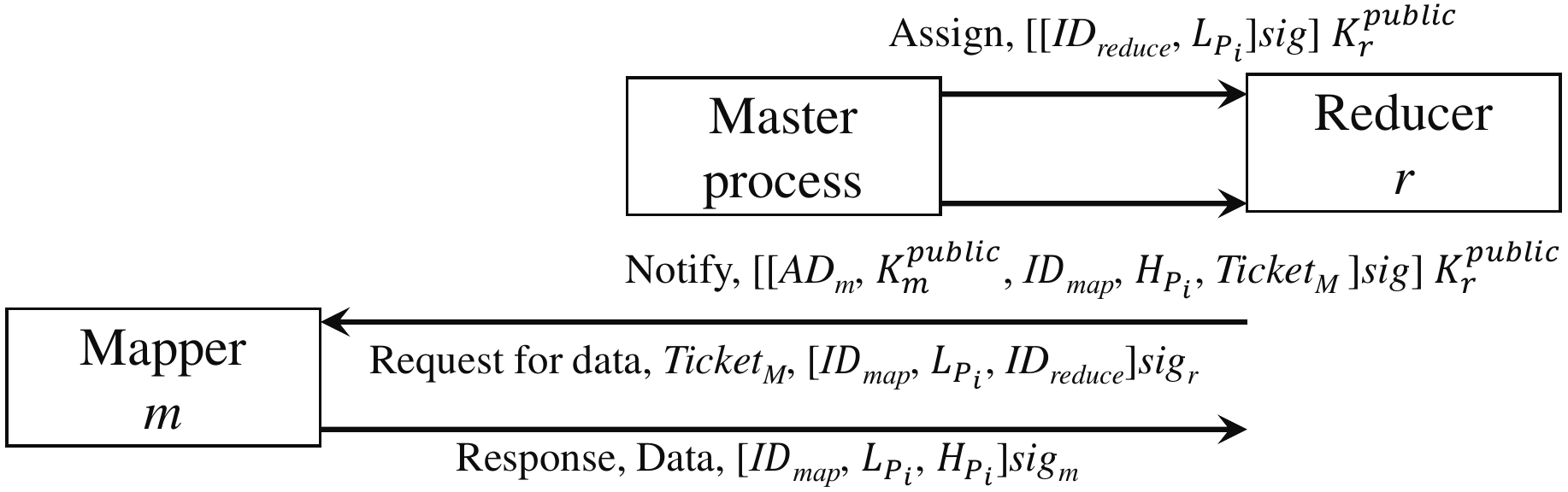}
                    \subcaption{Communication between the master process and a reducer, the verification protocol.}
                \label{fig:secure-mr-pic3}
            \end{minipage}
            \caption{SecureMR framework, the commitment protocol, and the verification protocol.}
\end{figure}

\medskip\noindent\textbf{SecureMR.} A decentralized replication-based integrity verification framework, called SecureMR~\cite{DBLP:conf/acsac/WeiDYG09}, ensures the integrity of data as well as computations, and prevents repudiation, DoS, and replay attacks. SecureMR replicates some map and reduce tasks, and assigns them to different mappers and reducers, \textit{i}.\textit{e}., a map (or reduce) task is executed by more than one worker. The proposed framework consists of five security components, as follows: Secure Manager, Secure Scheduler, Secure Task Executor, Secure Committer, and Secure Verifier; see Figure~\ref{fig:secure-mr-pic1}. The Secure Manager and the Secure Scheduler are deployed at the master process and perform task duplication, secure task assignment, and commitment-based consistency checking. The communication between the master process and mappers is carried out using the \emph{commitment protocol}. The communication between the master process and reducers, and between mappers and reducers is done using the \emph{verification protocol}.

\noindent \textit{Commitment protocol}. The commitment protocol, see Figure~\ref{fig:secure-mr-pic2}, avoids inspection of intermediate outputs by the master process and allows mappers to send a \emph{commit message} with a signed hash value for each of its intermediate outputs ($H_{P_{i}}$) to the master process. Note that in the commitment protocol, the master process assigns a map task to a mapper in a secure manner using an encrypted message (signed, $sig$, by the master process and encrypted using the public key of the mapper, $K^{public}_m$) that holds the location of data, $\mathit{Data_{Loc}}$, and identity of the map task, $\mathit{ID_{map}}$.

\noindent \textit{Verification protocol}. In the \emph{verification protocol}, see Figure~\ref{fig:secure-mr-pic3}, reducers verify intermediate outputs and check the signed hash value that was submitted to the master process. The master process assigns a reduce task to a reducer in a secure manner using an encrypted message (encrypted using the public key of the reducer, $K^{public}_r$) that holds the location of intermediate outputs, $L_{P_{i}}$, and identity of the reduce task, $\mathit{ID_{reduce}}$. After that the master process sends a \emph{notify message} to the verifier of each reducer, which includes the mapper's address ($\mathit{AD_M}$), $K^{public}_m$, $\mathit{ID_{Map}}$, $H_{P_{i}}$, and a ticket $\mathit{Ticket_M}$ (which includes $K^{public}_r$, $\mathit{ID_{Map}}$, $L_{P_{i}}$, and $\mathit{ID_{reduce}}$). The reducers ask intermediate outputs from the mapper, and the mapper sends data once it verifies the request. The verifier at the reducer verifies the response from the mapper, and in case of inconstancy, the verifier sends two signatures as evidences of an inconsistency to the master process.

\medskip In order to check the integrity of a MapReduce computation on a Desktop Grid~\cite{DBLP:conf/ccgrid/FedakGNC01,Anderson04boinc}, a replication based approach is suggested in~\cite{11desktopgridmrsecurity}, where reducers check results produced by mappers and the master process checks results produced by reducers. In addition, a MD5-based scheme is given to check outputs of mappers against a predefined digest code and Map function. The master process computes digest-codes for each split, and the digest-codes are sent to reducers. Mappers process input data splits, compute the code, and attach it with intermediate outputs. Reducers check the codes attached with intermediate outputs and the code received from the master process. If both the codes are different, then intermediate outputs are rejected.

\medskip\noindent\textbf{Overhead issues.} All the redundancy-based approaches replicate all or some of the tasks, and such a replication increases the communication cost and computation time of a MapReduce job. This becomes more important in non-free public clouds, where users are tariffed for the communication cost and computation time. An additional drawback of redundancy-based approaches is a difficulty to find a collusive malicious, as collusive attackers might provide an identical incorrect answer; thus bypassing the comparison-based verification.


\Subsubsubsection{Redundancy with trust based approaches}

Redundancy with trust based approaches use replication techniques for identification of inconsistencies, and then, use trusted workers for verification of results. Trusted workers might be located in a private cloud, in a different pool on the same cloud, or in any other trusted location. Also, it is assumed that the number of trusted workers is limited and does not allow execution of all tasks.

\medskip\noindent\textbf{Accountable MapReduce.} In order to ensure the integrity of MapReduce computations, Accountable MapReduce~\cite{DBLP:journals/fgcs/XiaoX14} verifies each map and reduce tasks by reexecuting them at a group of trusted workers. However, the reexecution of each task is not computationally efficient in detecting malicious mappers or reducers. Hence, the Accountable MapReduce also supports reexecution of some of the map and reduce tasks by a group of trusted workers.

\medskip\noindent\textbf{VIAF.} Verification-based Integrity Assurance Framework (VIAF)~\cite{DBLP:conf/IEEEcloud/WangW11} builds trust between the master process and each of the mappers, based on replication of tasks and a quiz-based system. The VIAF framework introduces a new task called \emph{verification task} that verifies outputs of the map tasks. The verification task, the reduce task, and the master process execute on a set of trusted workers; and the map task executes on untrusted workers. In the VIAF framework, each map task executes on two workers, and each worker accumulates a sufficient amount of credits by passing verification (quiz-based) process.

\begin{figure}[h]
 \centering
 \includegraphics[scale=0.45]{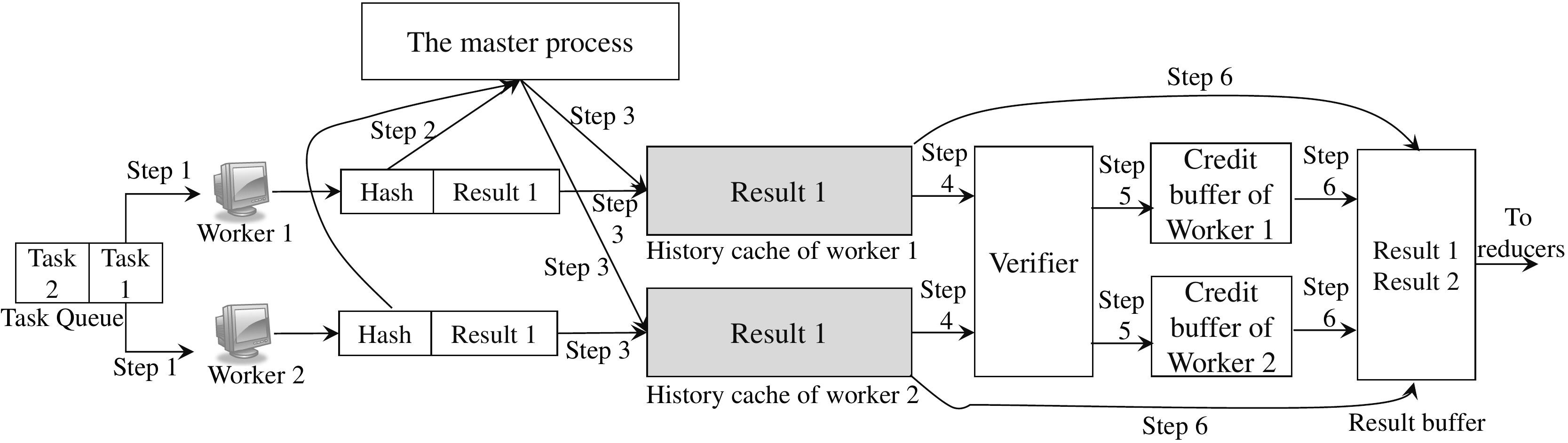}
 \caption{VIAF framework.}
 \label{fig:viaf-pic}
\end{figure}

In the VIAF framework, the master process assigns each map task to two different workers; see step 1 in Figure~\ref{fig:viaf-pic}. After completion of the map task, both the workers return hash values of the results to the master process; see step 2 in Figure~\ref{fig:viaf-pic}. If the hash values are different, then the master process concludes that one of them is a malicious (non-collusive) worker, and hence, assigns the same task to two different workers. On the other hand, if both the hash values are identical, the master process stores the results of the two workers and the task information in their \emph{history caches}; see step 3 in Figure~\ref{fig:viaf-pic}. After that, the master process may execute the verification task at the verifier in a non-deterministic manner (with a certain probability, called \textit{verification probability}) to verify these consistent results of both the workers; see step 4 in Figure~\ref{fig:viaf-pic}. The master process concludes that the two workers are collusive workers when the verification task provides different results. On receiving an identical result from the verification task, both the workers pass one quiz and accumulate one credit; see step 5 in Figure~\ref{fig:viaf-pic}. When both the workers have an adequate and an identical amount of credits, they become trusted workers, and their results are passed to reducers through a \textit{result buffer}; see steps 6 and 7 in Figure~\ref{fig:viaf-pic}.

\medskip\noindent\textbf{CCMR.} The framework, Cross Cloud MapReduce (CCMR)~\cite{ResultIntegritysinglepublic}, extends the VIAF~\cite{DBLP:conf/IEEEcloud/WangW11} framework by executing a MapReduce computation over a private cloud and a public cloud. The master process and the verification task are executed at the private cloud. In the map phase, each original map task is replicated with a replication probability, and each identical result is verified using the quiz-based approach with a verification probability. In the reduce phase, each reduce task is divided into multiple sub-tasks; and each sub-task is also replicated with a \textit{replication probability}, and each identical result of sub-tasks is verified using the quiz-based approach with a verification probability.

\medskip\noindent\textbf{IntegrityMR.} IntegrityMR~\cite{DBLP:conf/bigdataconf/WangWSDD13} extends the CCMR framework~\cite{ResultIntegritysinglepublic} by executing a MapReduce job on a private cloud and multiple public clouds, where the map and reduce tasks are executed on different public clouds. In addition, an invariant construction and checking
method for applications written in Pig Latin~\cite{Olston_piglatin} is also suggested, where an original script is transformed into another equivalent script. In the transformed script, a map task is substituted by two map tasks that work on some \emph{overlapped} inputs and provide an identical result as an invariant of the map task after processing overlapped inputs. In order to achieve high accuracy, the reduce task executes on the private cloud, detects invariant violation, and restores results from the two map tasks, if they provide identical results.

\medskip\noindent\textbf{VAWS.} Verification-based Anti-collusive Worker Scheduling (VAWS) system~\cite{DBLP:journals/ieicet/DingWWCFX14} improves the scalability of the VIAF framework by removing the bottleneck from the verifier and by enhancing the ability of the verification frameworks to deal with collusive attacks. Majority vote based systems have an inherent flaw when malicious attackers comprise a majority of computational nodes for specific work items. This situation might occur even if the amount of collusive malicious attackers in the cluster is small in comparison to the entire cluster size. The VAWS system deals with both challenges by separating reducers and the master process into trusted domain under the assumption that the majority of MapReduce jobs are mappers. Mapper jobs are executed on two nodes and the results are compared. Nodes that agree on results are considered to be temporary consistent. The algorithm then builds and maintains a sub-graph of consistent nodes. Nodes that disagree in outputs in any execution are then paired with other nodes, both consistent and not consistent, in order to identify malicious nodes. While the system has a low overhead of verification and experiments successfully identify malicious nodes, it has a big disadvantage of allowing rounds of computations with incorrect results. This happens when both nodes that execute an identical job are malicious and continue to happen until the malicious nodes are correctly identified.

\medskip\noindent\textbf{Hatman.} Another framework based on replication of tasks, Hatman (HAdoop Trust MANager)~\cite{DBLP:conf/IEEEcloud/KhanH12}, builds a trust level among different clouds, and NameNode keeps trust levels of DataNodes. A MapReduce job is distributed over $kn$ workers, where $k$ is a replication factor and $n$ is the size of a group that process an assigned job independently. In other words, $k$ non-identical groups of $n$ workers in each group process a job independently. If the system does not contain any malicious worker, then all the $k$ groups provide an identical result. On the other hand, when the master process receives different results, the master process chooses the results of trusted workers. Initially, all the nodes are assumed to be trusted and their relative confidence is also assumed to be uniform (\textit{i}.\textit{e}., $\frac{1}{n}$). However, workers decrease their trust and relative confidence in case of compromises. The trust of a worker $i$ towards a worker $j$ is proportional to the percentage of jobs shared by $i$ and $j$ on which $i$'s group agreed with $j$'s group, and a worker $i$ relative confidence is the percentage of assessments of $j$ that have been voiced by $i$.

\medskip\noindent\textbf{TrustMR.} In order to detect attacks with a high probability while minimizing the overhead, TrustMR~\cite{DBLP:conf/bigdataconf/UlusoyKP15} decomposes MapReduce tasks into smaller computations by means of aspect-oriented programming and replicates a subset of these task to verify the integrity of computations. TrustMR initiates multiple replicated map tasks on the replicated input splits. Some outputs of the map phase are randomly selected at runtime, and replicated map tasks only generate these key-value pairs. The results of replicated and original map tasks are verified at a map verifier by using a voting system. The results of replicated and original reduce tasks are also verified in the same manner at a reduce verifier.

\medskip\noindent\textbf{TS-TRV.} Trusted Sampling-based Third-party Result Verification (TS-TRV)~\cite{DBLP:dblp_conf/sose/DingWSFGZ13} performs random sampling and constructs a Merkle tree at a trusted third-party, called \textit{verifier}. The use of a Merkle tree reduces the amount of data that has to be sent to the verifier. A local Merkle tree is constructed by taking outputs of mappers as leaf nodes, when mappers finish the task. After that a global Merkle tree is constructed by the master process, where all root values submitted by mappers become leaf nodes, and the root value of the tree is sent to the verifier. The verifier randomly takes some number of (challenging) inputs from the global tree and sends them to corresponding mappers. On receiving (challenging) inputs, mappers construct responses in the form of a path to the root node of the local Merkle tree. The verifier also constructs the global Merkle tree using the responses sent by mappers and compares the values of the new root node against the old root node’s value (to find a malicious mapper).

\medskip\noindent\textbf{Overhead issues.} Most of the approaches check the trust of the workers before the execution of a MapReduce computation, which causes an overhead in addition to replication overhead. For instance, a malicious worker can behave well for a long period of time to gain the trust of the master process and may attack only after that. Most of the above approaches do not deal well with such sophisticated attackers. In addition, even replication of some tasks together with usage of trusted workers still cannot guarantee the detection of all the malicious mappers and reducers.


\Subsubsubsection{Log analysis and watermarking-based approaches}

In order to find malicious workers and a malicious update in HDFS, an approach based on log analysis is suggested in~\cite{CCGrid14loganalysis}. The approach verifies the integrity and the correctness of a MapReduce job without modifying the original MapReduce job. Four types of logs are recorded, as follows: (\textit{i}) logs of interaction between a user and NameNode and logs of interaction between a user and DataNode, (\textit{ii}) logs of HDFS access, (\textit{iii}) logs of interaction between HDFS and mappers-reducers, and (\textit{iv}) logs of the Map and the Reduce phases. These logs are compared with some pre-defined systems and job invariants to find malicious workers.

Watermarking uses the concept of a watermark that is a kind of indistinguishable marker embedded in data. An approach based on watermark (or probe) injection is given in~\cite{DBLP:conf/ccgrid/HuangZW12}, which is able to detect malicious and lazy (a worker process that can either drop a task at any time or start a task not from beginning) workers. The approach consists of four steps: \emph{watermark generation}, which generates some watermarks and inserts them into original data (before the start of a MapReduce job), \emph{execution of a MapReduce job}, \emph{verification}, which is done by the user by comparing outputs of all the processed watermarks by a MapReduce job and preprocessed outputs of all the watermarks, and \emph{recovery}, which removes injected watermarks from outputs, once the output passes the consistency check at the verification step. The approach works well mostly for text-intensive tasks, \textit{e}.\textit{g}., inverted index, word count, distributed grep, and log data processing, while being hard to apply to other types of input data. A method for verifying outputs of PageRank algorithm based on random sampling, called \emph{in-degree weighted sampling}, is also suggested in~\cite{DBLP:conf/ccgrid/HuangZW12}, which proposes a way for verifying outputs of tasks where watermark injection is not possible.

Another watermark based approach is suggested in~\cite{DBLP:conf/sose/DingWCTFS14}, which is also able to detect malicious workers. Accountable MapReduce~\cite{DBLP:journals/fgcs/XiaoX14} also supports watermarking-based detection of malicious workers. In Accountable MapReduce, some predefined watermarks are inserted into the original input data, and the outputs of the map phase, the inputs to the reduce phase, and the outputs of the reduce phase are verified.

A different approach was proposed and implemented in~\cite{7363786}. The paper suggested \emph{purpose-based access control (PBAC)} (see \cite{Byun:2005:PBA:1063979.1063998,Byun2006} for more details) and implemented it over Hadoop MapReduce. The set of purposes is organized in hierarchical tree, where an edge between two purposes represents relations (specialization and generalization) between them. The system modifies MapReduce jobs and records to include purposes. An access to specific data record is granted if the purposes specified by the security policy include or imply the purpose for the accessing of the data. It is assumed that when Hadoop system is deployed, the hierarchy of purposes is set in place and security policies are defined. A user submitting a job to Hadoop cluster then declares access policy, which is then checked by the AccountableMR system. The system clearly enhances the native security of Hadoop and allows much more fine-grained and sophisticated access control to the data. However, AccountableMR system also requires a considerable effort both in the initial setup of Hadoop cluster and from ongoing work of users.

\medskip\noindent\textbf{Restrictive issues.} Watermarking-based approaches are better than redundancy based approaches in terms of the workload on the framework, because these approaches do not re-execute all/some map and reduce tasks. However, watermarking approaches cannot be applied to all input data types or even to specific usages of the data. This considerably limits practical applications of such methods. In addition, the watermarking-based approaches do not guarantee finding all the malicious mappers and reducers, due to the fact that not all the data splits contain watermarks.

To summarize this section, there exist solutions to some of the security problems of MapReduce, however, more research is needed in order to provide effective solutions, which will be discussed briefly in Conclusions section (Section~\ref{sec:Conclusion}).

\section{Privacy Aspects in MapReduce}
\label{section:Privacy Aspects in MapReduce}
Privacy ensures that sensitive data is not exposed to untrusted users and trespassers (\textit{i}.\textit{e}., cloud providers, other data providers, users of MapReduce, or adversaries). Notice that the data providers are interested in allowing some sorts of computations on the data, however, there is also a requirement to preserve breach of sensitive data. Sensitive data in this case is case specific and might be personal records with identifier information (personally identifiable information PII), organization specific information and etc. In this section, we present a brief summary of privacy aspects in general cloud computing, privacy requirements in MapReduce, and then, review some existing solutions for privacy in MapReduce.

\subsection{Privacy Challenges in MapReduce Computing}
\label{subsection:Privacy Threats in MapReduce Computing}
Cloud computing and the deployment of MapReduce on public clouds present a new set of challenges in privacy of data. Here, we describe privacy challenges of cloud computing in the context of MapReduce and divide them into a few cases according to adversarial behaviors of public clouds and users.

\medskip\noindent\textbf{Data privacy protection from adversarial cloud providers.} A user may keep private data in public clouds due to its volume or ease of computations on public clouds, while aiming to preserve the privacy of data. In this setting, we consider an adversarial cloud provider that can observe users' data and MapReduce code; but should not change users' queries or results. Ensuring privacy in the presence of an adversarial cloud provider who can modify or delete data and computations is an insurmountable challenge. The goal of privacy in the presence of adversarial clouds is to minimize data leakage to the cloud provider while allowing users to perform operations on data. The majority of the work in this area is based on encryption of users' data and finding a way to allow operations on encrypted data in the cloud.

\medskip\noindent\textbf{Protection of data from adversarial users.} Data providers allow users to perform MapReduce jobs on their data via cloud providers, but also wish to control and preserve privacy of data. For example, while the average annual income can be calculated, the income of that specific individual should remain secret. Solutions to this use-case are based on anonymization of data (\textit{i}.\textit{e}., by dropping sensitive values that can identify individuals), by adding random noise to data, or computational results to hide the real values, (\textit{e}.\textit{g}., using differential privacy).

\medskip\noindent\textbf{Multiusers on a single public cloud.} A public cloud provider and a data provider should allow several users to perform their computations without data leakage. For example, an organization may keep all its data in public clouds, and several users process and access some parts of the data for which they are authorized. A hospital may store data of all the patients on clouds. There are several groups of users, \textit{e}.\textit{g}., doctors, insurance companies, patients, and pharmaceutical research companies, which should access some parts of data. In this case, privacy framework has to ensure that each user is able to access all the required data but also that the users cannot access parts of data for which they are not authorized for. This is usually solved by authentication and authorization mechanisms as was explained in Section~\ref{subsec:Proposed Solutions for Security in MapReduce}. The situation becomes more complex as data is provided by a number of data providers, each one with different privacy requirements. In such public clouds, an adversarial user may also access another user's data by injecting a malicious mapper or reducer that exploits existing security issues.

Various solutions are suggested for privacy of cloud computing. However, not all existing solutions for privacy of the cloud computing can be used for privacy in MapReduce, due to a number of additional constraints and challenges in MapReduce (presented in Section~\ref{sec:Challenges in MapReduce Environment and Adversary Models}), thus requiring adaptation or change in those solutions.

\subsection{Privacy Requirements in MapReduce}
\label{sebsec:Privacy Requirements in MapReduce}
MapReduce inherently decouples data providers, cloud providers, and users that execute queries over data. Referring to the cloud structure depicted in Figure~\ref{fig:pic_mr+cloud}, data providers upload data to the cloud provider, and cloud users perform queries on data. However, despite separation between different entities, ensuring privacy in those settings is still a challenging task. Here, we provide requirements of privacy in MapReduce framework, deployed on the hybrid cloud or the public cloud.

\medskip\noindent \textbf{Protection of data providers.} In a setting where data is uploaded to the cloud by various data providers, each data provider might have a different privacy requirements. The cloud provider has to ensure that those privacy requirements are met even in the presence of adversarial users. Moreover, different data providers might require a different privacy level for various data sets. The privacy framework should allow adaptation of privacy levels for those requirements.

\medskip\noindent\textbf{Untrusted cloud providers.} As an adversarial cloud provider can perform any computation on data for revealing data, modifying data, and producing wrong outputs, data has to be protected from cloud providers. In addition to protecting the data from cloud providers, privacy framework has to be able to protect the performed computations as well. As an example, consider a user querying for specific information. Even if the data results are not released to the cloud provider, it is possible to learn the intent of the user from observing performed computations. In terms of MapReduce, it may require mappers and reducers to work on encrypted data (see Section~\ref{subsec:Proposed Solutions for Privacy in MapReduce}).

\medskip\noindent\textbf{Utilization and privacy tradeoff.} A data provider can encrypt data in a way that no information can be learnt from it. However, this will also prevent the user from performing some computations on the data, and thus, decreases utilization of MapReduce. As such, MapReduce privacy framework has to provide maximum possible utilization while still preserving data privacy according to data providers' requirements.

\medskip\noindent\textbf{Efficiency.} In most of the public clouds, users are tariffed for usage and storage. Hence, the privacy framework has to be efficient in terms of CPU and memory consumption, and in the amount of storage required. If the privacy framework provides high overhead, it could be more cost-effective to perform computations on the private cloud, where physical security solves privacy issues.

\subsection{Adversarial Models for MapReduce Privacy}
\label{subsection:Adversary Models for MapReduce Privacy}
All the adversarial models mentioned in Section~\ref{sbsec:Adversary Models for MapReduce Security} are applicable to MapReduce privacy with small changes. Below, we explain the adaptation required in the definitions of adversaries and how they can be applied in privacy settings.

\medskip\noindent\textbf{Honest-but-Curious adversary.} This type of adversary mostly applies to cloud providers. Curious cloud providers can breach the privacy of data and MapReduce computations very easily, since the whole cluster is under the control of cloud providers, which have all types of privileged access to data and computing nodes. It is important to note that in reality curious cloud providers are not necessary adversaries by choice, but rather might be compliant by court law, regulations, and governmental requests.\footnote{http://www.zdnet.com/article/microsoft-admits-patriot-act-can-access-eu-based-cloud-data/}

\medskip\noindent\textbf{Malicious adversary.} This type of adversary applies to a user that tries to learn, modify, or delete information from the data by issuing various queries. In general, cloud providers are not assumed to be malicious, as assuring privacy with malicious cloud providers requires a high level of privacy measures that considerably reduce the utilization of the framework.

\medskip\noindent\textbf{Knowledgeable adversary.} A knowledgeable adversary applies to both a cloud provider and a user, who are trying to learn, modify, or delete information. Knowledgeable adversary is assumed to have a complete knowledge of MapReduce framework, the cloud structure, and is able to use any algorithm or cryptography drawback. In other words, there is no ``security by obscurity.''

\medskip\noindent\textbf{Network and node adversary.} As opposite to the network and nodes access adversary in security adversarial models, a cloud provider working as a network and node adversary has all the privileged access to computing nodes and the entire cloud infrastructure. A real-world example of such adversary is a cloud provider employee that breaches sensitive information most clearly shown by Edward Snowden case. It is impossible to hide any MapReduce computation or data from this type of adversary~\cite{insiderThreadStudy}.

\subsection{Proposed Solutions for Privacy in MapReduce}
\label{subsec:Proposed Solutions for Privacy in MapReduce}
This section summarizes some existing solutions for privacy in MapReduce. We categorize privacy algorithms in MapReduce into three types, as follows: (\textit{i}) algorithms for ensuring privacy in hybrid clouds, (\textit{ii}) algorithms ensuring data privacy in the presence of adversarial users, and (\textit{iii}) algorithms for ensuring privacy in the presence of adversarial cloud providers. A comparison of privacy algorithms, protocols, and frameworks for MapReduce is given in Table~\ref{table:Summary of privacy}.

\subsubsection{Data privacy in hybrid clouds}
\label{subsubsec:Data privacy in hybrid clouds}
An increasing growth of data within organizations and lower maintenance costs are two factors that force data processing on public clouds instead of private clouds. Despite the change in the location of data processing, the need for privacy preservation of sensitive data remains identical. Thus, it is beneficial to process data based on sensitivity on the organization's private cloud and public clouds. Since MapReduce is designed for a single cloud, hybrid cloud based MapReduce computations require modification of MapReduce framework in order to deal with privacy (and security) issues on public clouds. In this section, we review some MapReduce privacy frameworks for the hybrid cloud computing model.

\begin{figure}[h]
\begin{center}
    \begin{minipage}[t]{0.20\linewidth}
    \centering
    \includegraphics[scale=0.4]{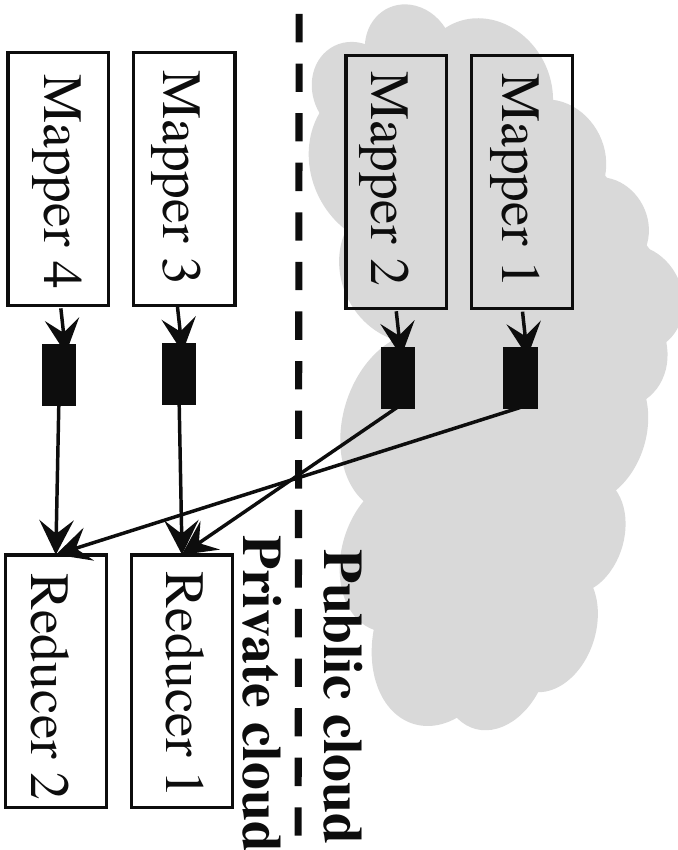}
    \subcaption{Map hybrid.}
    \label{fig:Map hybrid}
    \end{minipage}
    \begin{minipage}[t]{0.27\linewidth}
    \centering
    \includegraphics[scale=0.4]{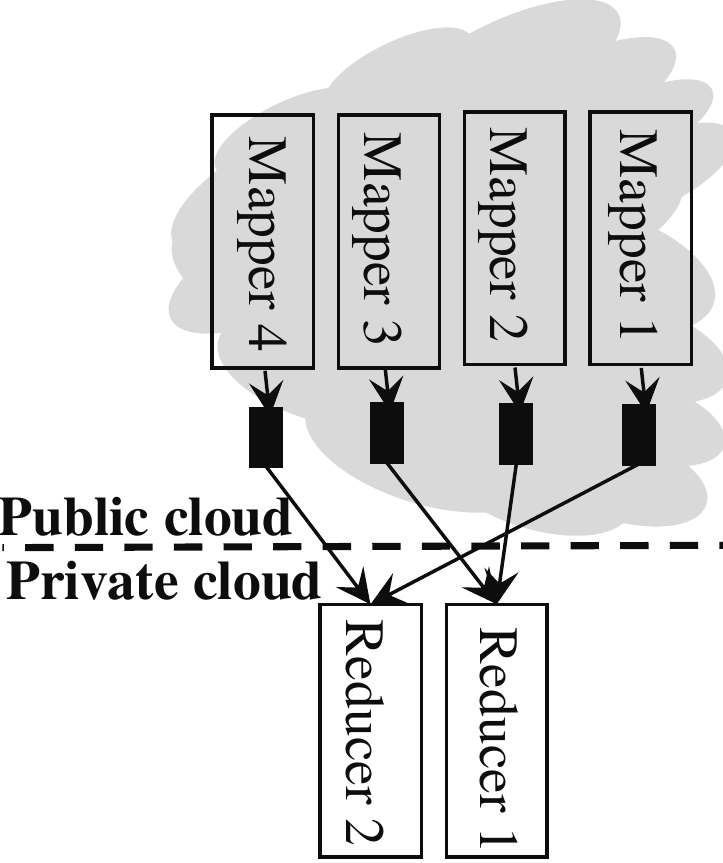}
    \subcaption{Horizontal partitioning.}
    \label{fig:Horizontal partitioning}
    \end{minipage}
    \begin{minipage}[t]{0.25\linewidth}
    \centering
    \includegraphics[scale=0.4]{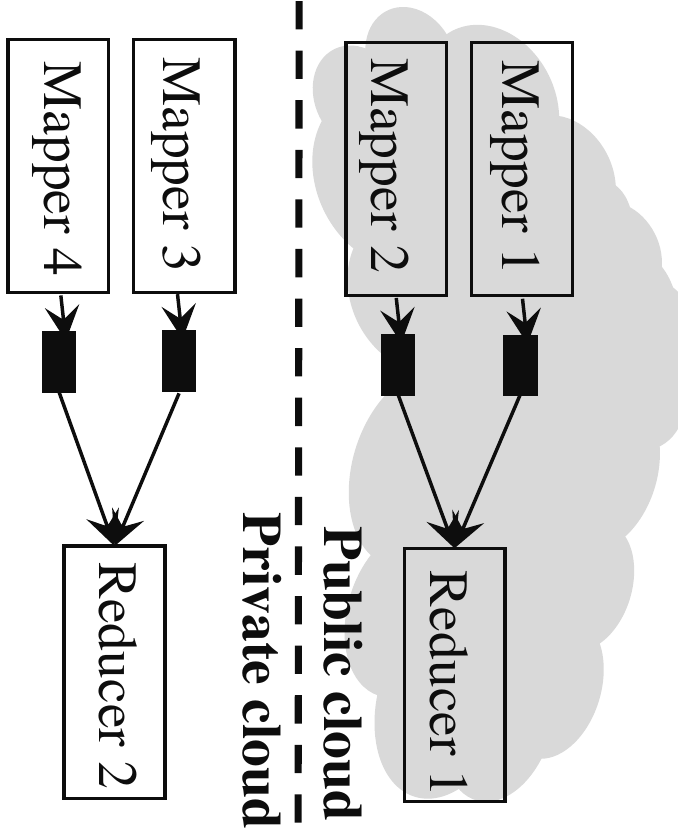}
    \subcaption{Vertical partitioning.}
    \label{fig:Vertical partitioning}
    \end{minipage}
    \begin{minipage}[t]{0.20\linewidth}
    \centering
    \includegraphics[scale=0.4]{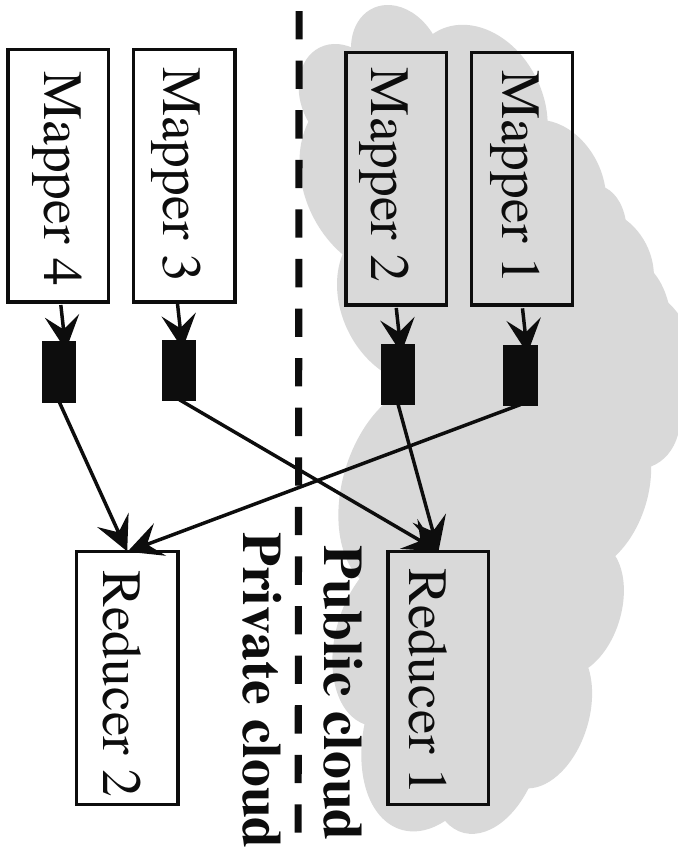}
    \subcaption{Hybrid.}
    \label{fig:Hybrid}
    \end{minipage}
\caption{Four execution models in HybrEx.}
\end{center}
\BB
\end{figure}

\medskip\noindent\textbf{HybrEx.} Hybrid Execution (HybrEx)~\cite{DBLP:conf/hotcloud/KoJM11} is the first MapReduce framework designed for the hybrid cloud. In HybrEx, data is divided into sensitive and non-sensitive data, non-sensitive data is sent to public clouds while sensitive data is kept in a private cloud. HybrEx allows four types of execution models of MapReduce computations, as follows: (\textit{i}) Map hybrid: the map phase is executed at both public and private clouds, however, the reduce phase is executed at a private cloud only (Figure~\ref{fig:Map hybrid}); (\textit{ii}) Horizontal partitioning: the map phase is executed (on encrypted data) at public clouds only, while the reduce phase is executed at a private cloud (Figure~\ref{fig:Horizontal partitioning}); (\textit{iii}) Vertical partitioning: the map phase and the reduce phase are executed on both public and private clouds while data transmission between private and public clouds is not allowed (Figure~\ref{fig:Vertical partitioning}); and (\textit{iv}) Hybrid: the map phase and the reduce phase are executed on both public and private clouds and data transmission among clouds is also possible (Figure~\ref{fig:Hybrid}). Two integrity check models, namely full integrity checking and quick integrity checking, are also suggested. However, HybridEx does not deal with a key that is generated at public and private clouds in the map phase.

\bgroup
\def\arraystretch{1.105}
\begin{table}
\begin{center}
\centering
\caption{Summary of privacy algorithms, protocols, and frameworks for MapReduce.}
\label{table:Summary of privacy}
\footnotesize
    \begin{tabular}{|p{3cm}|p{2cm}|l|l|l|p{1cm}|}
    \hline
    \multirow{2}{3cm}{Algorithms/Protocols/\\Frameworks} & \multirow{2}{2cm}{Privacy of data providers}& \multicolumn{2}{ c| }{Protection from adversarial} & Approach & \multirow{2}{1cm}{Cloud structure} \\

    \hhline{~~|-|-|~|}  & {~} & User  & Cloud &  &  \\ \hline

    HybrEx~\cite{DBLP:conf/hotcloud/KoJM11} & \checkmark & ~ & \checkmark & Data separation & H\tablefootnote{H: Hybrid cloud.} \\ \hline

    Sedic~\cite{DBLP:conf/ccs/ZhangZCWR11} & \checkmark & ~ & \checkmark & Data separation & H \\ \hline

    Tagged-MapReduce~\cite{DBLP:conf/ccgrid/ZhangCY14} & \checkmark & ~ & \checkmark & Data separation & H \\ \hline

    SEMROD~\cite{DBLP:conf/sigmod/OktayMKK15} & \checkmark & ~ & \checkmark & Data separation & H \\ \hline

    Prometheus~\cite{DBLP:conf/infocom/ZhouZDLY13} & \checkmark & ~ & \checkmark & Data separation & H \\ \hline

    PPL~\cite{DBLP:dblp_conf/cgc/ZhangLNDC12,zhang2014privacy} & ~ & \checkmark & ~ & Data anonymization & S\tablefootnote{S: Single cloud.} \\ \hline

   Airavat~\cite{DBLP:conf/nsdi/RoySKSW10} & \checkmark & \checkmark & ~ & Differential privacy & S \\ \hline

   PRISM~\cite{DBLP:conf/pet/BlassPMO12} & \checkmark & ~ & \checkmark & Encryption & S \\ \hline

PIRMAP~\cite{DBLP:conf/fc/MayberryBC13} & \checkmark & ~ & \checkmark & Encryption and PIR schema & S \\ \hline

EPiC~\cite{cryptoeprint:2012:452} & \checkmark & ~ & \checkmark & Homomorphic encryption & S \\ \hline

Powers and Chen~\cite{DBLP:journals/corr/abs-1211-3147} & ~ & \checkmark & ~ & Homomorphic Paillier encryption & S \\ \hline

PFC~\cite{DBLP:conf/IEEEcloud/XuSS14} & ~ & \checkmark & ~ & FPGAs and proxy re-encryption & M\tablefootnote{M: Multiple clouds.} \\ \hline

CryptDB~\cite{Popa11cryptdb} & \checkmark & ~ & \checkmark & Variable homomorphic encryptions & S \\ \hline

MrCrypt~\cite{DBLP:conf/oopsla/TetaliLMM13} & \checkmark & ~ & \checkmark & Variable homomorphic encryptions & M \\ \hline

Crypsis~\cite{crypsis} & \checkmark & ~ & \checkmark & Variable homomorphic encryptions & ~ \\ \hline

Dolev et al.~\cite{yinAAMR} & \checkmark & \checkmark & \checkmark & Secret-sharing & M \\ \hline
\end{tabular}
\BBB\BBB
\end{center}
\end{table}
\egroup

\medskip\noindent\textbf{Sedic.} In order to solve key problem of HybridEx~\cite{DBLP:conf/hotcloud/KoJM11}, Sedic~\cite{DBLP:conf/ccs/ZhangZCWR11} provides strategic data movement from the map phase that executes on public clouds to the reduce phase that executes on a private cloud, by using an automatic analysis and transformation of the reduce code. In order to decrease the communication between a public cloud and a private cloud, outputs of the map phase (at the public cloud) are aggregated before their transmission to the private cloud. In addition, Sedic framework automatically partitions a job by following security levels of data and distributes a job between private and public clouds.

\begin{figure}[h]
\begin{center}
    \begin{minipage}[t]{0.45\linewidth}
    \centering
    \includegraphics[scale=0.45]{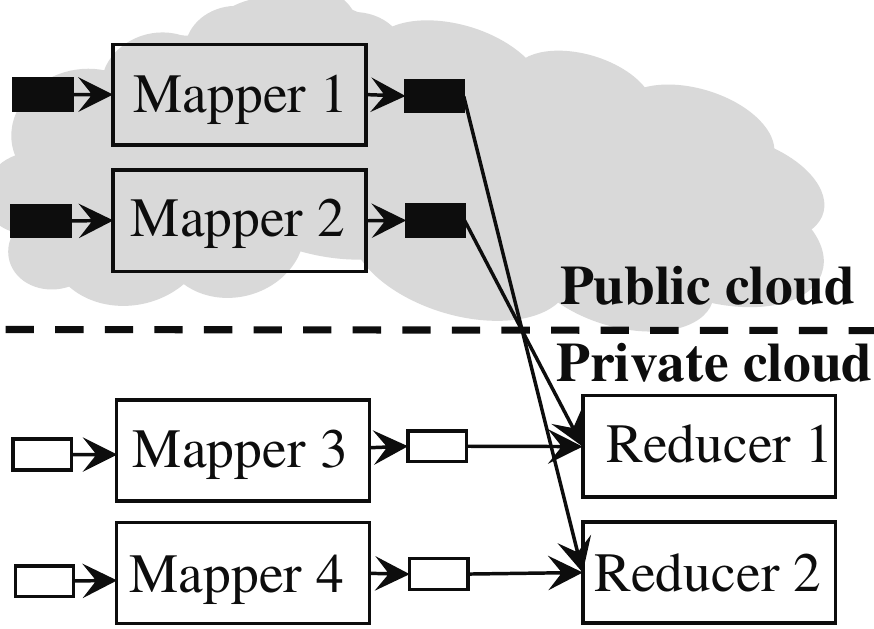}
    \subcaption{Single-phase mode.}
    \label{fig:pic_taggedmr1}
    \end{minipage}
    \begin{minipage}[t]{0.45\linewidth}
    \centering
    \includegraphics[scale=0.45]{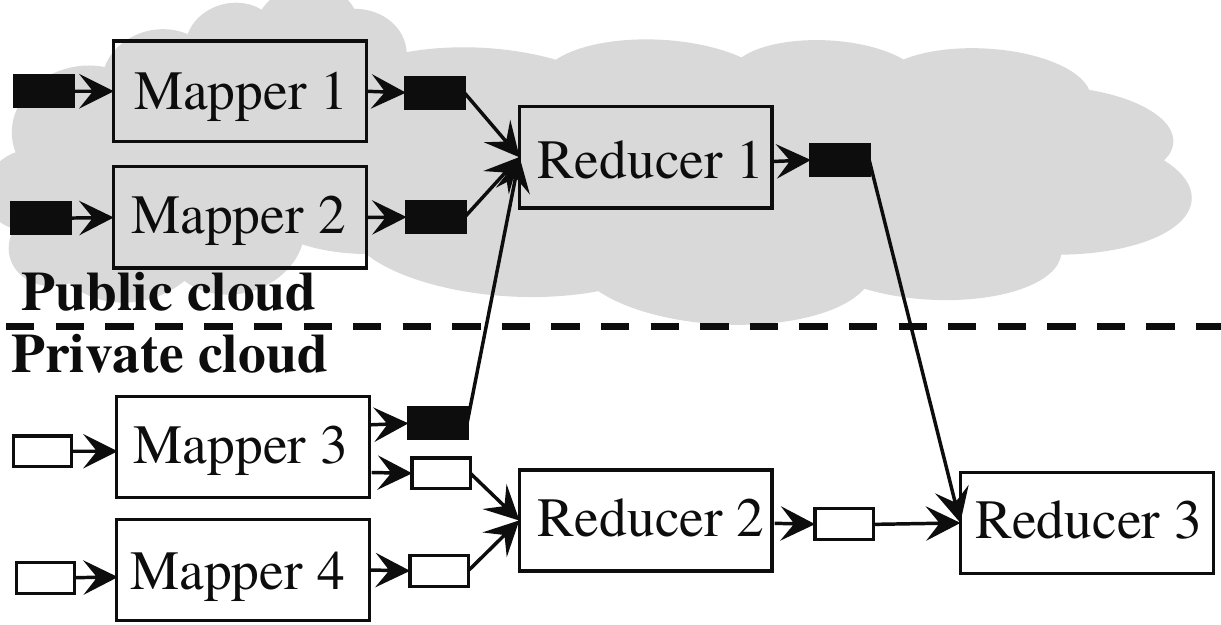}
    \subcaption{Two-phase crossing mode.}
    \label{fig:pic_taggedmr2}
    \end{minipage}
    \begin{minipage}[t]{0.35\linewidth}
    \centering
    \includegraphics[scale=0.45]{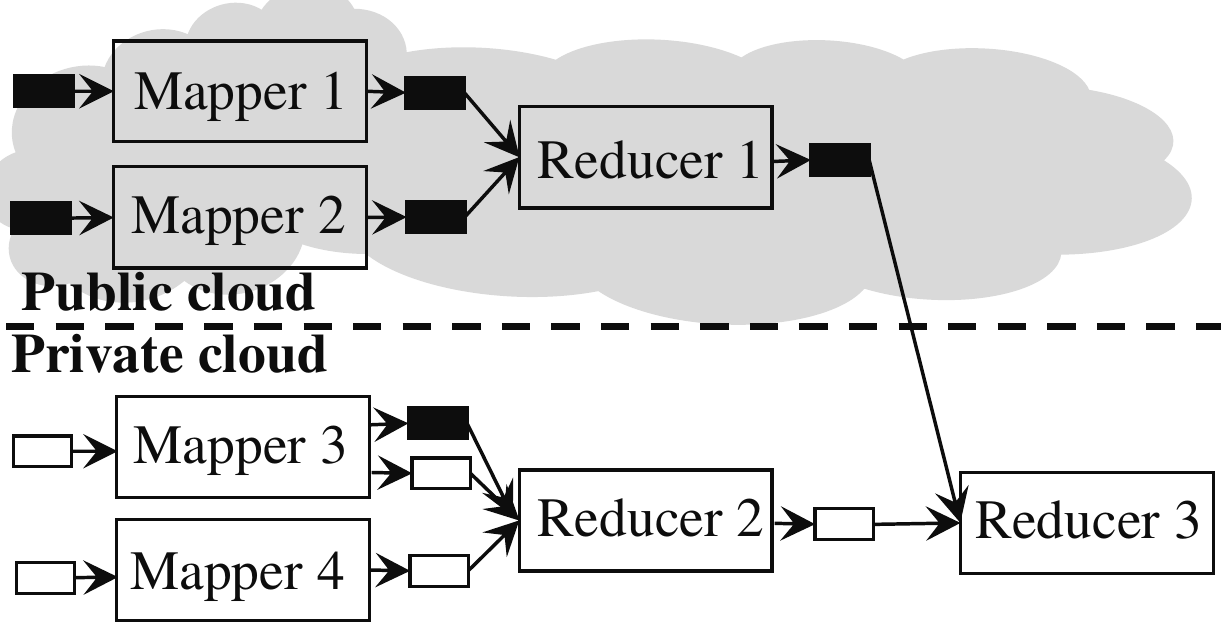}
    \subcaption{Two-phase non-crossing mode.}
    \label{fig:pic_taggedmr3}
    \end{minipage}
    \begin{minipage}[t]{0.35\linewidth}
    \centering
    \includegraphics[scale=0.45]{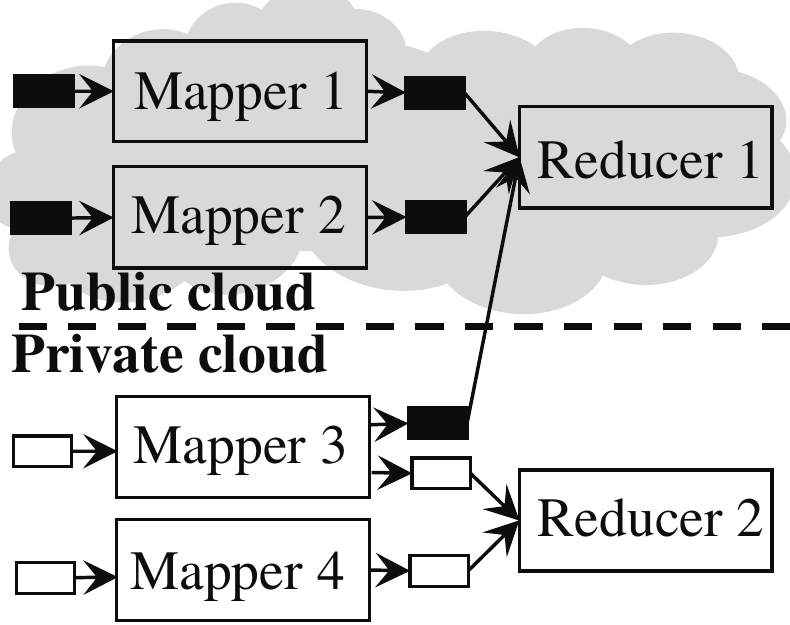}
    \subcaption{Hand-off modes.}
    \label{fig:pic_taggedmr4}
    \end{minipage}
   \begin{minipage}[t]{0.25\linewidth}
    \centering
    \includegraphics[scale=0.45]{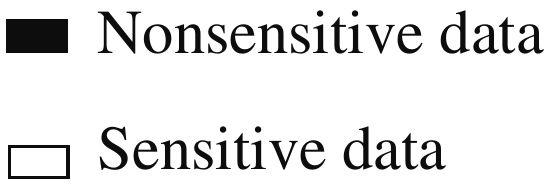}
    \subcaption{Notations.}
    \label{fig:pic_taggedmr5}
    \end{minipage}
\caption{Four scheduling modes in Tagged-MapReduce.}
\label{fig:tagged-mr}
\end{center}
\BB
\end{figure}

\medskip\noindent\textbf{Tagged-MapReduce.} HybrEx~\cite{DBLP:conf/hotcloud/KoJM11} and Sedic~\cite{DBLP:conf/ccs/ZhangZCWR11} consider data sensitivity before a job's execution. Tagged-MapReduce~\cite{DBLP:conf/ccgrid/ZhangCY14} identifies data-sensitivity during execution of a job, where the map phase and the reduce phase are executed on public and private clouds. The framework handles sensitivity of intermediate outputs that may contain sensitive data, and hence, cannot be processed by the reduce phase at public clouds. Two policies, non-upgrading policy and downgrading policy, help in identifying on-the-fly data sensitivity, and four scheduling modes (single-phase, two-phase crossing, two-phase non-crossing, and hand-off modes), see Figure~\ref{fig:tagged-mr}, assign outputs of the map phase to reducers regarding data sensitivity. In addition, Tagged-MapReduce supports iterative MapReduce jobs. However, HybrEx~\cite{DBLP:conf/hotcloud/KoJM11}, Sedic~\cite{DBLP:conf/ccs/ZhangZCWR11}, and Tagged-MapReduce~\cite{DBLP:conf/ccgrid/ZhangCY14} are unable to handle the situation efficiently when a key is generated at public and private clouds. In order to solve this, Oktay et al.~\cite{DBLP:conf/sigmod/OktayMKK15} suggested Secure and Efficient MapReduce Over hybriD clouds (SEMROD) that prevents the leakage of sensitive data and efficiently exploits public resources for executing a given single (or multi-level) MapReduce job.

\medskip\noindent\textbf{SEMROD.} SEMROD~\cite{DBLP:conf/sigmod/OktayMKK15} first finds sensitive and non-sensitive data and sends non-sensitive data to public clouds. Private and public clouds execute the map phase. However, instead of sending only outputs of the map phase containing sensitive keys to the private cloud, the private cloud pulls all the outputs, but executes the reduce phase operation only on record associated with sensitive keys and ignores non-sensitive keys. Public clouds execute the reduce phase on all the outputs. Hence, they are unable to know the sensitive keys. At the end, a filtering step removes duplicate entries, creating by sensitive key.

\medskip\noindent\textbf{Prometheus.} In order to outsource non-sensitive data, which is stored in relations, to public clouds, Prometheus~\cite{DBLP:conf/infocom/ZhouZDLY13} removes quasi-identifiers (a quasi-identifier refers to a subset of attributes that can uniquely identify most tuples in a relation~\cite{motwani2007qid,DBLP:conf/icde/PeiTLX09}) using a hypergraph. After the discovery of quasi-identifiers, attributes are distributed over public clouds, and an \emph{attribute location table} is used to store name of relations and the location of relations-attributes. This allows the system to ensure that no sensitive data is stored in untrusted public clouds. It also avoids heavy workload on reducers at the user-end by sending merged outputs of public clouds and a \emph{mapping table} of tuples from the private cloud to the user. Reducers (at the user-end) construct the final output. On the downside, Prometheus allows only search operations on a hybrid cloud.

\medskip A new framework for multiple clouds is proposed in~\cite{DBLP:journals/internet/LoughranCFKG12}. The framework is divided into three layers: (\textit{i}) \textit{the physical layer} -- holds computational resources; (\textit{ii}) \textit{the virtualization layer} -- allows users to share the computational resources in a secure and isolated manner; and (\textit{iii}) \textit{the infrastructure-as-a-service (IaaS) layer} -- manages and creates virtual resources, provides user management, and an access control method for accessing virtual resources. The IaaS layer has two sub-layers, namely \textit{automatic deployment layer} that (\textit{i}) creates virtual machines on the user-defined cloud, (\textit{ii}) installs and configures Hadoop's master process based on an assigned job, (\textit{iii}) executes a MapReduce job on the user-defined cloud, and \textit{monitoring layer} that monitors all the virtual machines and resources. The framework allows processing of sensitive data at a private cloud and processing of non-sensitive data at a public cloud. The framework uses existing encryption, authentication, and access control methods. A secure data exchange is carried out using secure transport protocols. After a job completes, data is deleted immediately from the virtual resources, virtual machines are cleaned, and results are sent back to the user. Hence, the cloud does not hold data for a long time, which incurs users to send data every time to the cloud before computations.

\medskip\noindent\textbf{Overhead issues.} In most of the cases, apart from Sedic~\cite{DBLP:conf/ccs/ZhangZCWR11}, the separation of sensitive and non-sensitive data is performed manually at a private cloud. Such a process drastically reduces the performance if datasets are huge. Moreover, a private cloud becomes a bottleneck if almost all the tuples contain sensitive data, which is processed on a private cloud.

\subsubsection{Data privacy with adversarial users}
\label{subsubsec:Data privacy with adversarial user}
In many applications, data providers and data users are different parties and might be completely separated. Examples of such applications are health service providers, pharmaceutical companies, and genomic data providers. In those cases, there is a clear need for providing data access to external parties for the purpose of research, monitoring, or knowledge sharing while providing sufficient data protection for data providers. However, for those purposes, the management of access controls or data encryption is not enough.

\emph{Data anonymization}~\cite{zhou2008brief,DBLP:journals/csur/FungWCY10} is a promising solution for ensuring data privacy on public clouds. It works by hiding data identifiers, \textit{i}.\textit{e}., attributes that allow identification of specific individuals, by changing information to some values, inserting records, and suppressing information~\cite{Adam89security-controlmethods,DBLP:conf/pods/MalvestutoMR91,CellSuppressionMethodology}.

Another notion of providing privacy, called \textit{Differential Privacy}~\cite{Dwork06differentialprivacy}. The idea of Differential Privacy is to ensure that an addition or removal of a single dataset item does not substantially affects the outcome of computations. In other words, adversary cannot distinguish between the results with and without a specific dataset item. The most common ways to achieve differential privacy is by addition of (specific) random noise to the sensitive data or computations, to hide an existence of any individual record. Differential Privacy is currently considered \textit{de facto} standard of private data publishing as it provides rather strong privacy guarantee as it does not depend on auxiliary information known to adversary or computational power.

Additional privacy preserving methods for MapReduce are encryption-decryption-based solutions~\cite{DBLP:conf/pet/BlassPMO12,DBLP:conf/fc/MayberryBC13} and an accountability-based solution~\cite{DBLP:journals/fgcs/XiaoX14}.

\begin{figure}[h]
 \centering
 \includegraphics[scale=0.45]{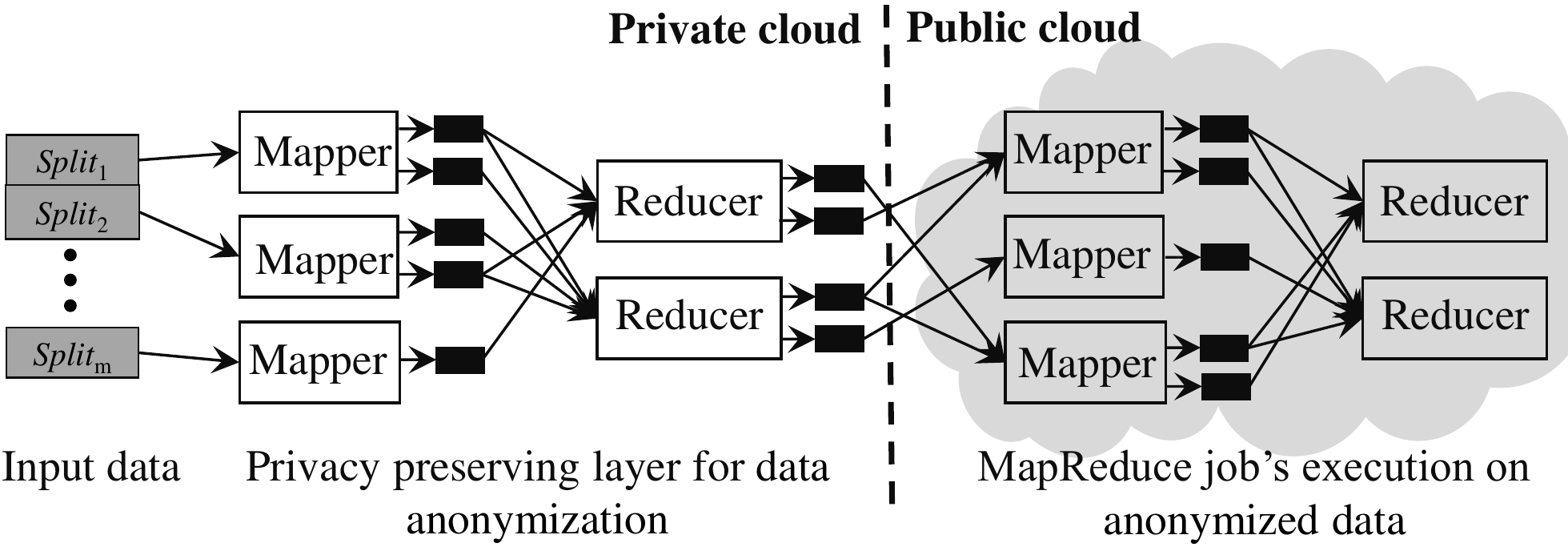}
 \caption{MapReduce framework with privacy-preserving layer.}
 \label{fig:ppl}
\end{figure}

In~\cite{DBLP:dblp_conf/cgc/ZhangLNDC12,zhang2014privacy}, the authors presented a new framework for MapReduce computations based on data anonymization. The framework introduces a new layer, called \textit{Privacy-Preserving Layer} (PPL, see Figure~\ref{fig:ppl}), that exists between the original data and MapReduce framework for executing an assigned job. The PPL layer takes privacy requirements and original data as inputs. The layer then can apply different anonymization approaches according to the privacy requirements of data providers. This allows flexibility in choosing different privacy mechanisms within a single framework.

Specifically, the PPL layer consists of four main modules, as follows: (\textit{i}) Privacy Specification Interface (PSI): takes several parameters as privacy specifications, (\textit{ii}) Data Anonymizing (DA): does anonymization of data using MapReduce based anonymization algorithms~\cite{NepalMRanon} and privacy specifications, (\textit{iii}) Data Update (DU): does anonymization of new data without anonymization of data from scratch, and (\textit{iv}) Anonymized Data sets Management (ADM): provides methods for storing anonymized data in public clouds without breaching privacy of data.

Airavat~\cite{DBLP:conf/nsdi/RoySKSW10} (see also Section~\ref{subsec:Access control based approaches}) is the first system that combines mandatory access control (MAC) and differential privacy for ensuring data privacy according to differential privacy definition from untrusted users. Airavat allows users to submit MapReduce jobs with custom mappers and reducers chosen from a pre-defined set of trusted reducers. Mandatory access control ensures that untrusted mappers do not leak data outside the system via network or file system. Differential privacy requirements are ensured on intermediate outputs by adding a random noise to them. In order to maximize the utilization of the system and minimize the amount of added noise, the user has to provide a range of output values for every provided mapper. If the output exceeds the range, it is re-mapped to a random value in the range. Naturally, the system requires full trust in the cloud provider that implements the protocol and ensures the integrity of its components.

\medskip\noindent\textbf{Utility issues.} All data anonymization methods have an inherent tradeoff between utilization and privacy. Since data anonymization methods provide a high level of data privacy, the system utilization may decrease. Also, it may be hard to ensure anonymized data after performing operations on it. For example, joining of a relations having anonymized data with other relation having non-anonymized data may reveal data of the first relation.\footnote{http://privacyguidance.com/blog/10-big-data-analytics-privacy-problems/}

\subsubsection{Data privacy in adversarial clouds}
\label{subsubsec:Data privacy in adversarial clouds}
The most obvious solution for ensuring privacy of data in public clouds is encryption of data; however, it creates hurdles for an efficient utilization of MapReduce. In this section, we present some existing techniques that enable cloud users to perform MapReduce computations on encrypted data, while preserving privacy of data.

\medskip\noindent\textbf{PRISM.} Privacy-Preserving Search in MapReduce (PRISM)~\cite{DBLP:conf/pet/BlassPMO12} alleviates the problem of storing data in curious cloud providers by allowing searching for any user specified word in privacy preserving manner, \textit{i}.\textit{e}., the cloud provider should not be able to learn the user query and data. The proposed protocol consists of three phases, as follows: (\textit{i}) upload of the data to the cloud, (\textit{ii}) search operation, and (\textit{iii}) result analysis phase. During the upload phase, the user encrypts data using state-full encryption algorithms, which add a frequency counters (as one of the possible options) to each word to prevent the cloud provider from computing statistics about frequency of encrypted text, and uploads the data to the cloud. In order to search the data, the user sends mappers and reducers based on Trapdoor Private Information Retrieval~\cite{trapdoorGroups2011} for acquiring search results. Note that the cloud provider is considered honest-but-curious, and it will not change received mappers and reducers.

\medskip\noindent\textbf{PIRMAP.} Another system leveraging Private Information Retrieval (PIR), first defined in~\cite{Chor:1995:PIR:795662.796270}, is Private Information Retrieval for MapReduce (PIRMAP)~\cite{DBLP:conf/fc/MayberryBC13}. PIRMAP is the first potentially practical cPIR algorithm (a cPIR algorithm is an algorithm that assumes that the cloud (or data) provider is polynomial-computational-bounded, as opposite to a generic case where the data provider is not bounded), which can be used in a real-world scenario. PIRMAP follows a ``classical'' PIR scheme (as defined~\cite{Chor:1995:PIR:795662.796270} and improved in~\cite{Lipmaa04anoblivious}) where the user sends an encrypted vector to the cloud provider. The cloud provider splits the data into blocks and multiplies each block by the received vector. Then, the cloud column-wise adds the results of the multiplication to create one-result vector. The vector is then returned to the user who decrypts it. These two stages of the algorithm, \textit{i}.\textit{e}., multiplication and column-wise sum, are quite easily mapped into two stages of MapReduce, \textit{i}.\textit{e}., map and reduce. Calculation of PIR scheme by MapReduce algorithm is done concurrently, according to the paradigm, and thus, allows great performance of otherwise computationally extensive scheme. The output of the mapper is a key-value pair, where the key is the index of the block and the value is a result of multiplication. Reducers then receive values of the column and perform the sum operation. Consequently, PIRMAP allows users to privately retrieve information from the cloud, using MapReduce.

\medskip\noindent\textbf{EPiC.} Efficient Privacy-Preserving Counting (EPiC)~\cite{cryptoeprint:2012:452} protocol allows privacy-preserving counting using MapReduce and allows users to store their data in public clouds privately, \textit{i}.\textit{e}. protected from curious cloud providers. At the first phase, the user encrypts the data and uploads it to the cloud. The data is encrypted in such way that an identical data value does not generate an identical ciphertext, and hence, the cloud provider, which stores the (encrypted) data, cannot learn anything from the data apart from trivial characteristics, such as data size. At the query stage, the user specifies a searching pattern as a Boolean formula and generates mapper/reducer code for working on the encrypted data. The computation is performed by using partially homomorphic encryption for protecting outputs of the computation from the cloud provider. The cloud provider performs an assigned MapReduce computation and counts the total number of occurrences of an assigned pattern without learning neither the data, the pattern, nor how often it occurs. EPiC is based on an idea of transforming the pattern search into a summation and polynomial evaluations, which can be done by partially homomorphic encryption scheme in an efficient manner. The protocol uses weaker encryption scheme for allowing more efficient execution of assigned queries. However, EPiC supports only counting operations, which is a limitation of the protocol.

A similar protocol was presented, in~\cite{DBLP:journals/corr/abs-1211-3147}, for allowing privacy preserving implementation of the power iteration algorithm (a method for finding dominant eigenvectors for large matrices) on MapReduce. The protocol uses partially homomorphic Paillier encryption~\cite{Paillier99public-keycryptosystems} scheme for algorithm computations. At the first stage of the protocol, the user encrypts the data using this encryption scheme and uploads the data to the public cloud. At processing stage, the user uses random vectors for protecting intermediate outputs and performs MapReduce computations by utilizing homomorphic properties of Paillier encryption scheme. The protocol is limited to computations of the specific algorithm only.

\medskip\noindent\textbf{PFC.} Field programmable gate arrays (FPGAs) and proxy re-encryption based privacy preserving solution for MapReduce computation is presented, in~\cite{DBLP:conf/IEEEcloud/XuSS14}, where data is kept in an encrypted form in public clouds. An encryption algorithm is selected in a manner that data is easily partitioned into a number of splits, to be processed by mappers. However, mappers and reducers are not allowed to process encrypted splits and intermediate outputs, respectively. Mappers decrypt assigned splits before processing them and again encrypt intermediate outputs. The reducer also first decrypt intermediate outputs before processing and decrypt final outputs.

\medskip\noindent\textbf{CryptDB.} CryptDB~\cite{Popa11cryptdb} executes SQL queries over encrypted data providing practical confidentiality for the users. The idea of CryptDB is that most of the queries use well-defined set of operations, each of which is possible to support efficiently over encrypted data. CryptDB protects data from curious DBA that snoops on the database server and from a curious cloud provider that holds the servers and the data. The adversary does not change user queries. The tradeoff is between strong encryption, which will not allow many operations on the data, and between weaker encryption with more operations. Another tradeoff is minimizing the amount of leaked data when application servers are compromised. The authors do not see arbitrary computations on encrypted data as practical; thus, the application server has to be able to process decrypted data. (Their analysis over 128,840 queries from MIT applications showed that CryptDB can support 99.5\% of all queries. It reduces throughput by 14.5\% for full Web forums and by 26\% for TPC-C queries comparing to unmodified MySQL.)

\medskip\noindent\textbf{MrCrypt.} Following the work of CryptDB~\cite{Popa11cryptdb}, MrCrypt~\cite{DBLP:conf/oopsla/TetaliLMM13} suggests a way for executing MapReduce computations on encrypted data stored on curious cloud providers. MrCrypt's privacy preserving mechanism is based on two observations, as: (\textit{i}) many MapReduce jobs perform only a limited set of basic operations on input data and (\textit{ii}) homomorphic encryption schemas that enable specific operations are much more efficient than fully homomorphic encryption (\cite{Gentry:2009:FHE:1536414.1536440,Gentry11Implementing11}).

\begin{figure}[h]
\centering
\includegraphics[scale=0.45]{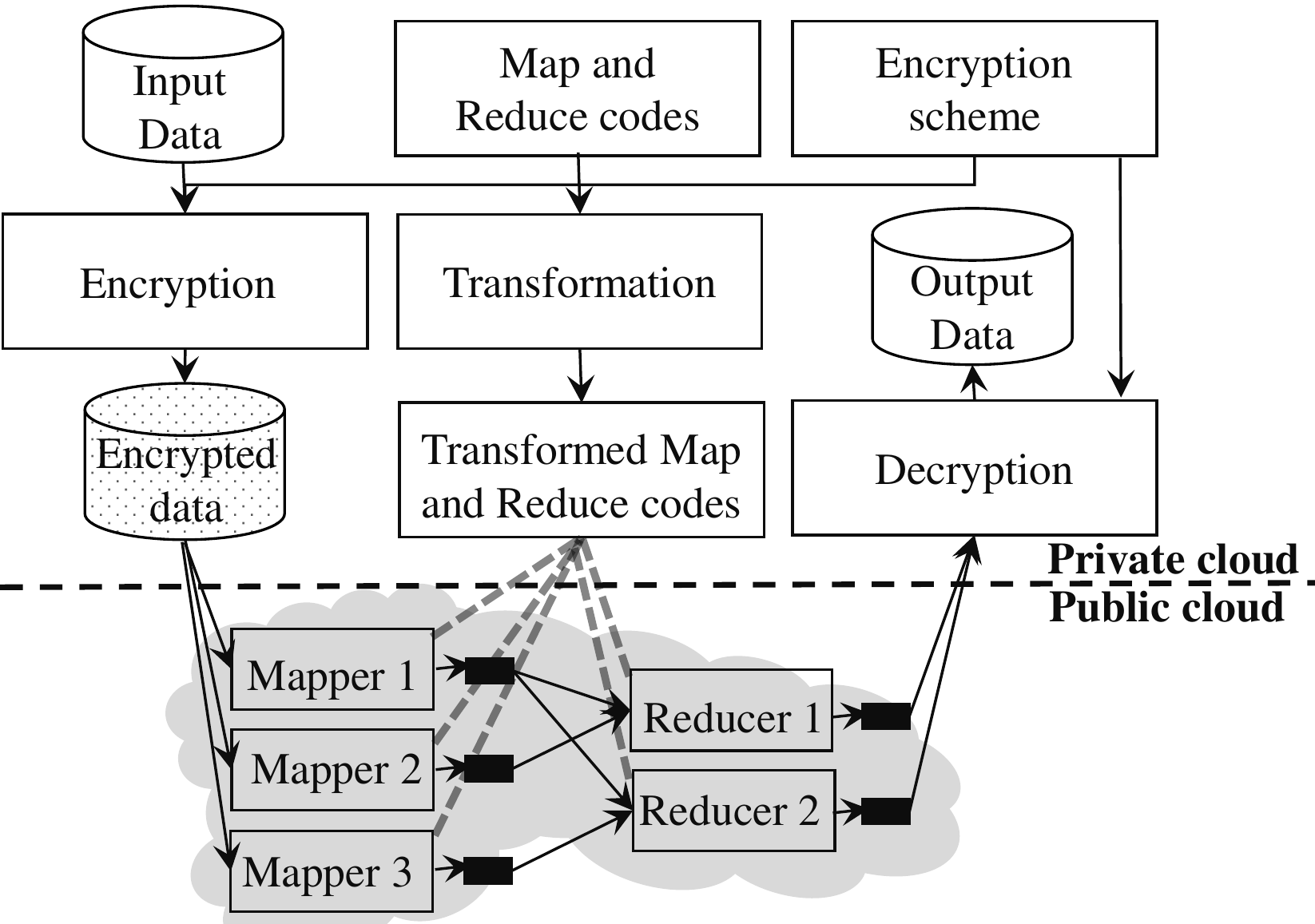}
\caption{MrCrypt framework.}
\label{fig:mrcrypt}
\end{figure}

MrCrypt, see Figure~\ref{fig:mrcrypt}, performs static analysis of Java code for mapper and reducers at a private cloud. Following the analysis, a minimal homomorphic encryption scheme is chosen for supporting all the required operations in a legal and correct manner. The Java programs are then transformed using this encryption scheme, and data is also encrypted using the scheme. Next, the user uploads data and the transformed programs to a public cloud provider, which executes MapReduce job. The final outputs of the job are sent back to the user and are decrypted using the chosen homomorphic scheme. However, the downside of the approach is that it limits the range of possible queries on the system.

\medskip\noindent\textbf{Crypsis.} The ideas of CryptDB~\cite{Popa11cryptdb} and MrCrypt~\cite{DBLP:conf/oopsla/TetaliLMM13} were taken to higher data languages by Crypsis~\cite{crypsis}. The system enables execution of Pig Latin jobs on a curious cloud provider without exposing data. Crypsis executes a MapReduce job on encrypted data without decrypting it. In order to do that, the system transforms a Pig Latin script so that it can be executed on encrypted data. Crypsis uses existing practical partially homomorphic encryption schemes for data encryption. The system works in the following phases: (\textit{i}) script transformation, Pig Latin script is analyzed and required encryption schemes are identified, the script is then changed to use encrypted data; (\textit{ii}) update cloud with missing encryption schemes: it is possible that data stored in the cloud is missing some encryption schemes that are required for the given script, in that case those schemes are identified and the cloud is updated with newly encrypted data; (\textit{iii}) execute encrypted script on the cloud infrastructure using pre-defined code provided by user stored with the data; (\textit{iv}) re-encryption, it is possible that intermediate outputs are generated during the execution of the script, in such cases the data should be re-encrypted and (\textit{v}) results, the results are sent to the user where they can be decrypted.

\medskip\noindent \textbf{Overhead issues.} The main obstacle of providing privacy-preserving framework for MapReduce cloud computations with a cloud provider acting as an adversary is computational and storage efficiency. The currently known fully homomorphic encryption schemes computational overhead is still prohibitively expensive~\cite{Gentry11Implementing11}; thus, there is a need to find new schemes or methods of ensuring data privacy. The research papers reviewed in this section show that a considerable advance was made towards this goal. Nevertheless, all the above mentioned algorithms have common drawbacks such as: limited range of allowed queries (as a tradeoff between preserving data privacy and utilization), increased computation time, and in many cases an increased storage space for storing encrypted data. Despite those difficulties, the future of privacy preserving computations in public clouds looks promising and interesting.

\subsubsection{Data privacy in adversarial clouds using secret-sharing}
\label{subsubsec:Data privacy in adversarial clouds using secret-sharing}
All the approaches suggested in Section~\ref{subsubsec:Data privacy in adversarial clouds} are based on encryption-decryption, which comes at a
price of computation and limited operations~\cite{DBLP:conf/isi/UlusoyKTK15}. In~\cite{yinAAMR}, a Shamir's secret-sharing~\cite{DBLP:journals/cacm/Shamir79} based solution for five types of MapReduce computations such as count, search, fetch, equijoin, and range selection is provided. The creation of secret-shares is computationally less expensive as compared to encryption-decryption techniques, and it provides information-theoretically secure computation. The suggested approach makes secret-shares of data and sends them to non-communicating clouds. A user can execute MapReduce computations of the form of secret-shares in those clouds and receives an answer of the form of secret-shares. By performing an interpolation technique, the user get the desired result. By using secret-sharing of data and computation, the cloud cannot learn the database and computation. Also, the user cannot learn the whole database. However, the use of more than one cloud increases costs.

\section{Conclusions and Future Research Directions}
\label{sec:Conclusion}
Processing a huge amount of data is not simple using the classical parallel computing, due to the failure of computing nodes and scalability of the system. MapReduce, developed by Google in 2004, provides an efficient, fault tolerant, scalable, and transparent processing of large-scale data. However, MapReduce was not designed to be deployed on public and hybrid clouds, where security and privacy of data and computations are two prime concerns. Since public clouds provide an easy way for computations and storage, a number of algorithms and frameworks regarding security and privacy of data-computations were developed for executing a MapReduce job on public and hybrid clouds.

In this survey, we discussed security and privacy challenges and requirements in MapReduce. Security attacks in MapReduce -- impersonation, denial-of-services, replay, eavesdropping, man-in-the-middle, and repudiation attacks -- are presented. We consider four types of adversarial models, namely honest-but-curious, malicious, knowledgeable, and network and nodes access adversaries, and show how they can impact a MapReduce computation. We reviewed many of the existing algorithms and frameworks for ensuring security and privacy in the scope of MapReduce.

Existing algorithms and frameworks succeed in solving the specific security and privacy problems in MapReduce. For example, data transmission and data storage are protected by encryption mechanisms; authentication and authorization solutions are based on existing secret key and integrated systems (such as SELinux); the result verification is done by replication of tasks; and privacy is ensured using data anonymization, differential privacy, and private information retrieval. Privacy preserving research is still struggling with providing a high utilization of MapReduce framework. While the reviewed papers show potentially practical solutions for specific problems, there is still considerable overhead (in terms of the workload on the framework) and limitations in utilization of the framework.

Based on this survey, we identified several important issues and challenges that require further research, as follows:
\begin{itemize}
\item Extending the authorization framework (security of MapReduce), \textit{i}.\textit{e}., how to incorporate advanced authorization policies (\textit{e}.\textit{g}., role-based or attribute-based access control policy) in MapReduce framework? This is particularly important if the mappers need to access different sources of data within the cluster.

\item Integrating with a trust infrastructure (security of MapReduce). There are several domains of trust that must be made explicit and verified for MapReduce framework. These include: trust in the hardware, virtual machine, and file system that mappers and reducers use, trust that MapReduce code is not malicious or does not try to leak confidential data, and trust in the cloud provider for providing the necessary resources to run MapReduce algorithms.

\item Processing on encrypted data (security and privacy of MapReduce). Although, as we have seen, some work has been done in this area, especially using homomorphic encryption, more research is needed in order to enable various MapReduce algorithms on encrypted data.

\item Supporting multiple geographically distributed clusters for executing a single job (security and privacy of MapReduce). Often data and computing resources for a single job may exist in different independent clusters. For example, a bio-informatic application that tries to analyze genomes existing in different countries and labs in order to track the sources of a potential epidemic. How MapReduce can be extended to multiple clusters, with support for privacy of the sensitive information across the clusters is an open research problem.

\item Extending MapReduce algorithms with privacy preserving support (privacy of MapReduce). These include support for secure computations between reducers and across clusters. Also, privacy policies, which define exactly what kind of aggregated or anonymized data can be output, need to be defined.
\end{itemize}

Another lacking field of the research is holistic frameworks (with a salient exception of Airavat~\cite{DBLP:conf/nsdi/RoySKSW10}), \textit{i}.\textit{e}., frameworks that solve more than a single problem, especially solving both the security and privacy aspects, and integrating some of the mentioned algorithms and frameworks, which provide computational security and privacy of data for MapReduce computations. We believe that in the future we will have MapReduce frameworks that provide multiple types of computations in information secure manner.

\phantomsection
\section*{Acknowledgements}
Part of this research was supported by a grant from EMC Corp. We thank the project coordinator Dr. Patricia Florissi for this support and very helpful comments.


\end{document}